\documentclass[12pt, draftclsnofoot, onecolumn]{IEEEtran}
\ifCLASSINFOpdf

\else

\fi
\usepackage[cmex10]{amsmath}
\usepackage{amssymb}
\usepackage{cite}
\usepackage{graphicx}
\usepackage{array,color}
\usepackage{multirow}
\usepackage{amsmath}
\usepackage{stfloats}
\usepackage{graphicx}
\usepackage{subfigure}
\usepackage{tabularx}
\usepackage{epsfig,epsf,color,balance,cite}
\usepackage{setspace}
\usepackage{bm}
\usepackage{textcomp}
\usepackage[linesnumbered, ruled]{algorithm2e}
\usepackage{subfigure}
\usepackage{caption}
\usepackage{graphicx}
\usepackage{setspace}
\SetKwRepeat{Do}{do}{while}%

\begin{document}

\title{Resource Allocation for Secure URLLC in Mission-Critical IoT Scenario}

\author{
Hong Ren, Cunhua Pan, Yansha Deng, Maged Elkashlan, and Arumugam Nallanathan, \IEEEmembership{Fellow, IEEE}
\thanks{H. Ren, C. Pan, M. Elkashlan, and A. Nallanathan are with  Queen Mary University of London, London, E1 4NS, U.K. (Email: {h.ren, c.pan, maged.elkashlan, a.nallanathan}@qmul.ac.uk). Y. Deng is with King's College London, London WC2R 2LS, U.K. (e-mail: yansha.deng@kcl.ac.uk).}
}

\maketitle

\vspace{-0.5cm}
\begin{abstract}
Ultra-reliable low latency communication (URLLC) is one of three primary use cases in the fifth-generation (5G) networks, and its research is still in its infancy due to its stringent and conflicting requirements in terms of extremely high reliability and low latency. To reduce latency, the channel blocklength for packet transmission is finite, which incurs transmission rate degradation and higher decoding error probability. In this case, conventional resource allocation based on Shannon capacity achieved with infinite blocklength codes is not optimal. Security is another critical issue in mission-critical internet of things (IoT) communications, and physical-layer security is a promising technique that can ensure the confidentiality for wireless communications as no additional channel uses are needed for the key exchange as in the conventional upper-layer cryptography method.
This paper is the first work to study the resource allocation for a secure mission-critical IoT communication system with URLLC. Specifically, we adopt the security capacity formula under finite blocklength and consider two optimization problems: weighted throughput maximization problem and total transmit power minimization problem. Each optimization problem is non-convex and challenging to solve, and we develop efficient methods to solve each optimization problem. Simulation results confirm the fast convergence speed of our proposed algorithm and demonstrate the performance advantages over the existing benchmark algorithms.
\end{abstract}

%

%
\newpage

\section{Introduction}

The fifth-generation (5G) networks are expected to support three main use cases: enhanced mobile broadband (eMBB),  massive machine type communication (mMTC), and ultra-reliable low latency communication (URLLC) \cite{Shafi2017}. Significant advancement has been achieved in the last decade for the use case of eMBB characterized by high throughput and data rate. Some typical techniques include massive multiple-input multiple-output (MIMO) and mmWave communications. For the use case of mMTC, 5G networks aim to provide massive connectivity to tens of billions of low-cost small-size machine-type devices such as smart glasses, smart thermometers, wireless sensors, etc. Some access protocols such as random access and grant-free access are shown to be effective in mMTC. However, the realization of URLLC is more challenging than eMBB and mMTC due to the fact that URLLC targets at two stringent quality-of-service (QoS) requirements in term of extremely high reliability (e.g., $1-10^{-9}$) and ultra-low latency (e.g., 1 ms), which are conflicting with each other. Specifically, to achieve high reliability, long codeword with redundancy is required, which increases the latency. On the other hand, short packet/codeword is mandated to achieve low latency, which lowers the reliability performance. The research of URLLC is still in its infancy and is main target of Release 17. In addition, URLLC is closely relevant to mission-critical internet of things (IoT) applications with emphasis on high reliability and low latency, such as autonomous factory manufacture, remote surgery, unmanned aerial vehicles (UAVs) control, and vehicular communication networks.

The primary feature associated with URLLC compared with conventional human-to-human communications is its short packet transmission feature, which is adopted to guarantee ultra-low latency. In this case, the law of large numbers is not valid and Shannon capacity cannot be applied to characterize the system capacity. Knowing that short blocklength is adopted in URLLC, the decoding error probability will not approach zero even when the signal-to-noise ratio (SNR) is arbitrarily high. If Shannon capacity expression is directly applied for transmission design, the reliability and latency will be underestimated, and the QoS cannot be guaranteed. In \cite{Polyanskiy2010}, the authors first derived the approximate expression of the data rate for a point-to-point AWGN channel under the case of finite channel blocklength, which is a function of the SNR, channel blocklength, and decoding error probability. Recently, this information-theoretical result has been adopted to design the resource allocation in various communication systems, e.g., simultaneous wireless information and power transfer (SWIPT) system in \cite{chen2019resource}, unmanned aerial vehicle (UAV) communications in \cite{pan2019joint,hongren2019wcl}, non-orthogonal multiple access (NOMA) in \cite{sun2018short}, mobile edge computing in \cite{she2019cross} and  factory automation scenario in \cite{hong-twc,hong-jasc,changyang2018}.

On the other hand, due to the broadcast nature of wireless communications, IoT applications such as industrial robots are particularly vulnerable to security threats (e.g. critical control information leakage or malicious attack). Conventionally, the security is enhanced through cryptography at the upper layers of the communication system. However, the secret key exchange and management is complicated and needs additional channel uses to accomplish these protocols. In URLLC, channel blocklength is limited, and the cryptography method may not be applicable in URLLC applications. On the other hand, physical layer security, which exploits the nature of wireless channels, is more favourable for URLLC as the complicated key exchange procedure is unnecessary. Recently, physical-layer security has been extensively studied in existing literature \cite{mukherjee2014principles,mukherjee2015physical,li2019joint,yizhoutcom}. However, infinite blocklength is assumed in these papers, and the security capacity is defined as the highest coding rate that there always exists a pair of channel encoder and decoder such that both the decoding error probability at the legitimate receiver and the information leakage to the eavesdropper can be made arbitrarily small when the channel blocklength is sufficiently large.

Unfortunately, the security capacity formula based on the infinite blocklength assumption is not applicable for secure URLLC applications, where short channel code/blocklength is adopted to reduce the latency. There is only a limited number of contributions studying the secrecy rate under finite blocklength \cite{weiyang-tit-19,huimingwang2019}. Most recently, the approximate security capacity formula under finite blocklength has been derived in \cite{weiyang-tit-19}, which is more complicated than the simple point-to-point communication system in \cite{Polyanskiy2010}. Based on this information-theoretical result, the authors in \cite{huimingwang2019} analyzed the performance of secure short-packet communications in a mission-critical IoT system with an eavesdropper. However, the resource allocation based on this result has not yet been studied.

Against the above background, the resource allocation problem for a secure mission-critical IoT system under short packet communications is studied in this paper. Specifically, the contributions are summarized as follows:
\begin{enumerate}
  \item We first consider the weighted sum throughput (WST) maximization problem by jointly optimizing the bandwidth unit and power allocation, while guaranteeing the total power and bandwidth constraints. This optimization problem is challenging to solve due to the following reasons. First, this problem involves the discrete variables associated with the number of bandwidth unit allocation. Second, the optimization variables are coupled in the objective function. Hence, this problem is a non-convex mixed-integer programming. To handle this problem, we develop an efficient iterative algorithm based on the principles of block coordinate descent (BCD) along with the successive convex approximation (SCA) method to solve this problem. Both the convergence and complexity analysis are provided. Greedy search method is adopted to convert the continuous variables into discrete ones.
  \item We then jointly optimize the power and channel bandwidth unit allocation to minimize the total transmit power (TTP) for a mission-critical IoT system under short packet communications, while guaranteeing the minimum security capacity of each device and the total channel blocklength. The optimization problem is a non-convex and mixed-integer programming problem, and NP-hard to solve. We first express the power for each device as a function of channel blocklength, and relax the discrete constraint for the channel blocklength to continuous variables. Then, a sufficient condition is proposed when the channel blocklength allocation problem is a convex optimization problem. This condition holds for typical URLLC application scenarios. At last, greedy method is used to convert the continuous solutions to discrete solutions.
  \item Finally, simulation results confirm the performance advantage of the proposed algorithm over the benchmark solutions such as the conventional long packet transmission scheme, which verifies the importance of adopting the security capacity formula under finite channel blocklength in the system design.
\end{enumerate}

The rest of this paper is organized as follows. System model and problem  formulation are provided in Section \ref{jnohtoi}. Weighted sum throughput maximization problem is solved in Section \ref{througput}, while total transmit power minimization problem is considered in Section \ref{ttpmin}. Simulation results along with related discussions are shown in Section \ref{simul}. Finally, conclusions of this paper  are drawn in Section \ref{jjitrj}.

\IEEEpeerreviewmaketitle

\section{System Model and Problem Formulation}\label{jnohtoi}
\subsection{System model}
\begin{figure}
\centering
\includegraphics[width=4.1in]{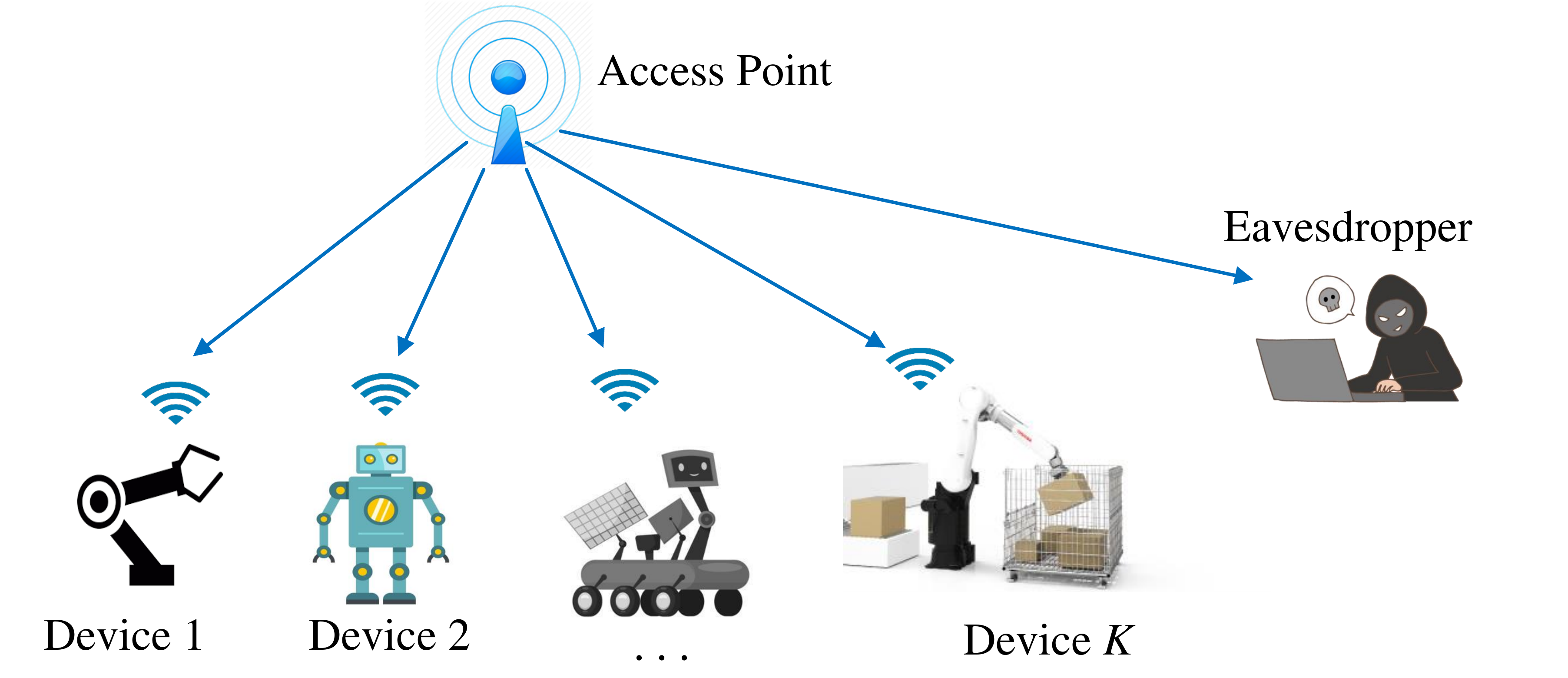}
\caption{Illustration of a secure mission-critical IoT communication system.}
\vspace{-0.1cm}
\label{systemodel}
\end{figure}
We consider a downlink mission-critical IoT communication system as depicted in Fig. \ref{systemodel}, in which an access point (AP) transmits confidential control signals to $K$ wireless connected devices (e.g. actuators, robots, and automated guided vehicle (AGV)). Meanwhile, there is an eavesdropper that aims to intercept the critical control signals transmitted by the AP. The AP, $K$ devices, and the eavesdropper are assumed to be equipped with a single antenna. Due to low-latency transmission, it is not feasible to allocate different time slots to all devices. Instead, we assume that all the devices are allocated with orthogonal frequency bands while transmitting over the same time duration, denoted as $T$.

In practical systems, the frequency band is divided into multiple basic bandwidth units with bandwidth $B_0$. Each device is assumed to operate in different frequency bands and the total frequency bandwidth allocated to the $k$th device is denoted as $B_k=n_kB_0$, where $n_k$ denotes the number of bandwidth units allocated to the $k$th device. We assume that the total bandwidth allocated to all the devices should be no larger than channel
coherence bandwidth $W_c$. It is assumed that $W_c$ is divisible by $B_0$, which can be expressed as $W_c=n_{\rm{max}}B_0$. Therefore, we have
\begin{equation}\label{kypju}
 \sum\nolimits_{k = 1}^K {{n_k}}  \le n_{\rm{max}}.
\end{equation}
Then, the number of channel uses allocated for the $k$th device is given by $B_kT$. In URLLC, the transmission duration $T$ is extremely small, which is shorter than the channel coherence time. Hence, the channels from the AP to $K$ devices and the eavesdropper  stay constant over each transmission. The channels  from the AP to the $k$th device and the eavesdropper are denoted as $h_k^d\in \mathbb{C}$ and  ${{h}}^e\in \mathbb{C}$, respectively.
Then, the received signal to noise ratio (SNR) of the $k$th device is given by
\begin{equation}\label{ijfjojtgr}
\gamma^d_k= \frac{{{p_k}{g_k^d}}}{{{n_k}}},
\end{equation}
where $p_k$ is the transmit power for the $k$th device, and $g_k^d={|h_k^d|^2}/({\sigma_{d,k}^2}B_0)$ with ${\sigma_{d,k}^2}$ denoting the noise power  spectrum density at the $k$th
device. It is assumed that the eavesdropper can access to all the frequency bands occupied by the devices. Thus, when the eavesdropper attempts to eavesdrop the $k$th device's information,  the received SNR at the eavesdropper is given by
\begin{equation}\label{ijjrjs}
\gamma^e_k= \frac{{{p_k}{g^e}}}{{{n_k}}},
\end{equation}
where $g^e={{| {{{{h}}^e}}|^2} \mathord{\left/
 {\vphantom {{| {{{{h}}^e}}|^2} {\left( {\sigma _{e}^2{B_0}} \right)}}} \right.
 \kern-\nulldelimiterspace} {\left( {\sigma _{e}^2{B_0}} \right)}}$ and  ${\sigma_{e}^2}$ is the noise power  spectrum density at the eavesdropper.

\subsection{Achievable Secrecy Data Rate Under Finite Blocklength}

It is well-known that when the number of channel uses is sufficiently large and the transmission data rate is lower than the secrecy capacity, we can always find a channel coding scheme such that both the decoding error probability and information leakage can be made as small as possible. In URLLC, the transmission blocklength (or the number of channel uses) is finite to guarantee low latency. However, short blocklength transmission suffers from a non-zero decoding error probability and non-negligible information leakage.

Based on \cite{weiyang-tit-19}, for a given channel blocklength $N_k=B_kT$, to guarantee a maximum decoding error probability of  $\epsilon_k$ at the $k$th device, and a secrecy constraint on the information leakage of $\delta_k$,  a lower bound on maximum secrecy communication rate (bit/s/Hz) can be approximated by:
\begin{equation}
\label{reteexpression}
r_k={C_k}-\sqrt{\frac{V^d_k}{N_k}}\frac{Q^{-1}(\epsilon_k)}{\ln2}-\sqrt{\frac{V^e_k}{N_k}}\frac{Q^{-1}(\delta_k)}{\ln2}
\end{equation}
where ${C_k}=\log_2(1+\gamma^d_k)-\log_2(1+\gamma^e_k)$ denotes the maximum secrecy capacity that can be achieved under infinite channel blocklength,  $V_k^x=1-(1+\gamma_k^x)^{-2}$, $x \in \{d,e\}$ is the channel dispersion which characterizes the random variability of a channel with respect to a deterministic channel with the same capacity \cite{hongrenICC}, and $Q^{-1}(\cdot)$ is the inverse of the Q-function  $Q(x)=\int^\infty_x\frac{1}{\sqrt{2\pi}}e^{-\frac{t^2}{2}} dt$. Similar to conventional physical layer security systems, $\gamma_k^d>\gamma_k^e$ should hold to ensure a positive data rate, which is equivalent to ${g_k^d}>{g^e}$ according to (\ref{ijfjojtgr}) and (\ref{ijjrjs}). The total number of bits (or throughput) that can be transmitted for each transmission for the $k$th device is given by
\begin{eqnarray}
   \! R_k &=&  n_kB_0Tr_k \\
   &=&  {n_k}{B_0}T \left(  {  C_k-\sqrt {\frac{{V_k^d}}{{{n_k}{B_0}T}}} \frac{{{Q^{ - 1}}({\epsilon_k})}}{{\ln 2}}  - \!\sqrt {\frac{{V_k^e}}{{{n_k}{B_0}T}}} \frac{{{Q^{ - 1}}({\delta _k})}}{{\ln 2}}} \!\!\right).\label{gedwjij}
\end{eqnarray}

In the following two sections, we aim to jointly optimize the number of bandwidth units and the power allocation to maximize the weighted sum throughput (WST) and minimize the total transmit power (TTP), respectively.

\section{Weighted Sum Throughput Maximization}\label{througput}

In this section, we aim to maximize the WST of all devices through jointly optimizing the  number of bandwidth units and the power allocation. Specifically, we first provide the problem formulation. Then, one efficient algorithm  is proposed to solve the optimization problem.

\subsection{Problem Formulation}

For the case when the AP places more emphasis on the amount of information transmitted, we aim to jointly optimize the power allocation and the bandwidth unit  allocation to maximize the WST of all devices while guaranteeing the total power constraint at the AP, and the total number of available bandwidth units. Thus, the optimization problem can be formulated as follows:
\begin{subequations}
\begin{align}
({\bf{P1}}):\;\;\max_{{\bm{p}},{\bm{n}}}\;\;\;& \sum\nolimits_{k=1}^K {\omega _k}R_k \\
\text{s.t.}\;\;\;\;\;\;&\sum\nolimits_{k=1}^{K} p_k\le P_{\max},\label{powerlimit} \\
& \sum\nolimits_{k=1}^Kn_k\le n_{\rm{max}},\label{blocklength} \\
&n_k\in \mathbb{N}^+, \;\;\forall k=1,\cdots, K,\label{blocklength-cons}\\
&p_k\ge 0, \forall k=1,\cdots, K, \label{nonegp}
\end{align}
\end{subequations}
where ${\bm{p}}=\{p_1,\cdots,p_K\}$, ${\bm{n}}=\{n_1,\cdots, n_K\}$, ${\omega _k}$ is a positive weight factor used to ensure the fairness among the devices and $\mathbb{N}^+$ denotes the non-negative integer set. Inequalities (\ref{powerlimit}) and (\ref{blocklength}) correspond to the total power constraint and total bandwidth constraint, respectively. Constraint (\ref{blocklength-cons}) means that the integer constraint for the number of bandwidth units. Note that $R_k=0$ when $p_k=0$, which ensures the non-negative value of $R_k$ in the optimal solution.

Problem $(\textbf{P1})$ is a mixed integer programming problem due to the non-negative integer constraints on $\bm{n}$. To make it tractable, we relax the integer $\bm{n}$ to continuous variables, and then convert the continuous solutions into integer ones. Therefore, Problem $(\textbf{P1})$ is relaxed as follows:
\begin{subequations}
\begin{align}
({\bf{P2}}):\;\;\max_{{\bm{p}},{\bm{n}}}\;\;\;& \sum\nolimits_{k=1}^K \omega _kR_k \\
\text{s.t.}\;\;\;\;\;\;&(\ref{powerlimit}), (\ref{blocklength}), (\ref{nonegp}), \\
&n_k\ge 0, \;\;\forall k=1,\cdots, K.\label{blockldfh}
\end{align}
\end{subequations}
However,  Problem $\textbf{(P2)}$ is still difficult to solve since ${\bm{p}}$ and ${\bm{n}}$ are coupled together. To circumvent this difficulty,  we adopt the block coordinate descent (BCD) to decouple these optimization variables. In particular, we optimize one set of variables while keeping the others fixed, and vice versa. Then, each subproblem is solved in an iterative manner. Specifically, Problem $\textbf{(P2)}$  is decoupled into two subproblems as
\begin{subequations}
\begin{align}
&({\bf{P2-1}}):\;\; \max_{\bm{p}} \;\;\;\;\sum\nolimits_{k=1}^K  \omega _kR_k(p_k)\;\;\;\;\;\; \text{s.t.}\;\;(\ref{powerlimit}), (\ref{nonegp}) \nonumber \label{p1-a}\\
&({\bf{P2-2}}):\;\; \max_{{\bm{n}}} \;\;\;\;\sum\nolimits_{k=1}^K  \omega _kR_k(n_k)\;\;\;\;\;\; \text{s.t.}\;\;(\ref{blocklength}), (\ref{blockldfh})\nonumber
\end{align}
\end{subequations}
where Problem $({\bf{P2-1}})$ corresponds to the optimization of power allocation ${\bm{p}}$ with given channel blocklength allocation ${\bm{n}}$, while Problem $({\bf{P2-2}})$ is the optimization of channel blocklength ${\bm{n}}$ with given ${\bm{p}}$. Each subproblem will be solved in the following subsections.

\subsection{The Solution of $\text{(P2-1)}$}

In this subsection, we aim to solve the power allocation of Problem $({\bf{P2-1}})$ with given $\bm{n}$. To this end, we first define $\bar g_k^d \buildrel \Delta \over = \frac{{g_k^d}}{{{n_k}}}$, $\bar g_k^e \buildrel \Delta \over = \frac{{g^e}}{{{n_k}}}$, ${L_k^d} = \frac{{{Q^{ - 1}}({\epsilon_k})\sqrt {{N_k}}}}{{\ln 2 }}$, and ${L_k^e} = \frac{{{Q^{ - 1}({\delta_k})\sqrt {{N_k}}}}}{{\ln 2 }}$. Then, $R_k(p_k)$ can be rewritten as
\begin{equation}\label{dewgt}
  {R_k}({p_k}) = \underbrace {{N_k}{{\log }_2}\left( {1 + {p_k}\bar g_k^d} \right) - {N_k}{{\log }_2}\left( {1 + {p_k}\bar g_k^e} \right)}_{f_k({p_k})} - \underbrace {\left( {\sqrt {V_k^d} {L_k^d} + \sqrt {V_k^e} {L_k^e}} \right)}_{y_k({p_k})}.
\end{equation}
Since $g_k^d>g^e$, we have $\bar g_k^d>\bar g_k^e$.

Before solving Problem $({\bf{P2-1}})$, we first analyze  the concave-convex property of $R_k(p_k)$ with respect to (w.r.t.) $p_k$. To this end, we first show that both functions ${f_k({p_k})}$ and ${y_k({p_k})}$ are concave w.r.t. $p_k$ as proved in the following lemma.

\emph{\bf{Lemma 1}}:  ${f_k({p_k})}$ and ${y_k({p_k})}$ are concave w.r.t. $p_k$, and thus $R_k(p_k)$ is the difference of two concave functions $f_k(p_k)$ and $y_k(p_k)$.

\emph{Proof}: Please refer to Appendix \ref{lemma1}.   \hfill\rule{2.7mm}{2.7mm}

According to Lemma 1, both ${f_k({p_k})}$ and ${y_k({p_k})}$ are concave w.r.t. $p_k$. Hence, for given $\bm{n}$, the objective function
of Problem $({\bf{P2-1}})$ is a difference of two concave functions, and thus is a non-concave function w.r.t. $p_k$. As a reslut, Problem $({\bf{P2-1}})$ is a non-convex optimization problem, and the globally optimal solution is difficult to find. However, Problem $({\bf{P2-1}})$ belongs to a class of difference of convex (DC) problems \cite{dinh2010local}, where the objective is to maximize a difference of two concave functions. This type of optimization problem can be efficiently solved by using the   successive convex approximation (SCA) method, which solves the DC problem in an iterative manner.

Denote the solution of $\bm{p}$ in the $(i-1)$-th iteration as ${\bm{p}}^{(i-1)}$. By exploiting the concavity of $y_k(p_k)$ and Jensen's inequality, we have
\begin{equation}\label{dfretg}
 y_k({p_k}) \le y_k\left( {p_k^{(i - 1)}} \right) + \beta _k(p_k^{(i - 1)})\left( {{p_k} - p_k^{(i - 1)}} \right),
\end{equation}
where $\beta _k(p_k^{(i - 1)})$ is the first-order derivative of $y({p_k})$ at $p_k^{(i - 1)}$, and is given by
\begin{equation}\label{xdesfrvgt}
  \beta _k(p_k^{(i - 1)}) = \frac{{{{\left( {1 + p_k^{(i - 1)}\bar g_k^d} \right)}^{ - 3}}\bar g_k^d{L_k^d}}}{{{{\left( {1 - {{\left( {1 + p_k^{(i - 1)}\bar g_k^d} \right)}^{ - 2}}} \right)}^{\frac{1}{2}}}}} + \frac{{{{\left( {1 + p_k^{(i - 1)}\bar g_k^e} \right)}^{ - 3}}\bar g_k^e{L_k^e}}}{{{{\left( {1 - {{\left( {1 + p_k^{(i - 1)}\bar g_k^e} \right)}^{ - 2}}} \right)}^{\frac{1}{2}}}}}>0.
\end{equation}

By replacing $y_k({p_k})$ with  the right hand side (RHS) of (\ref{xdesfrvgt}), we obtain the optimization problem to be solved in the $i$th iteration, which is given by:
\begin{subequations}
\begin{align}
({\bf{P2-1-a}}): \;\;\max_{{\bm{p}}}\;\;\;& \sum\nolimits_{k = 1}^K {\left( \omega _k{f_k\left( {{p_k}} \right) - \omega _k{\beta _k}(p_k^{(i - 1)}){p_k}} \right)}  \\
\text{s.t.}\;\;\;\;\;\;&(\ref{powerlimit}), (\ref{nonegp}),
\end{align}
\end{subequations}
where the constant values have been omitted in the objective function. Note that the objective function of Problem $({\bf{P2-1-a}})$ is a concave function and its constraints are affine functions of $\bm{p}$. Then, Problem $({\bf{P2-1-a}})$ is a convex optimization problem. The optimal solution of Problem $({\bf{P2-1-a}})$ can be found in the following theorem.

\emph{\bf{Theorem 1}}:  The optimal solution of Problem $({\bf{P2-1-a}})$ is given by
\begin{equation}\label{defrfe}
  {p_k^\star}(\lambda) = {\left[ {\frac{{ - \left( {\bar g_k^d + \bar g_k^e} \right) + \sqrt {{{\left( {\bar g_k^d + \bar g_k^e} \right)}^2} - 4\bar g_k^d\bar g_k^e\left( {1 - \eta _k^{(i - 1)}}(\lambda) \right)} }}{{2\bar g_k^d\bar g_k^e}}} \right]^ + },\forall k
\end{equation}
where $[x]^+$  is equal to ${\rm{max}}\{x,0\}$ and $\eta _k^{(i - 1)}(\lambda)$ is given by
\begin{equation}\label{xscd}
  \eta _k^{(i - 1)}(\lambda) = \frac{{{N_k}}}{{\ln 2}}\frac{\omega _k({\bar g_k^d - \bar g_k^e})}{{\omega _k{\beta _k}(p_k^{(i - 1)}) + \lambda }}, \forall k.
\end{equation}
If ${\sum\nolimits_{k = 1}^K {p_k^ \star (0 )} \le {P_{\max }}} $, then $\lambda=0$. Otherwise,  $\lambda$ is the root of the following equation:
\begin{equation}\label{sdefrgef}
   {\sum\limits_{k = 1}^K {p_k^ \star (\lambda )}  - {P_{\max }}} = 0.
\end{equation}

\emph{Proof}: Please refer to Appendix \ref{Theorem1}.   \hfill\rule{2.7mm}{2.7mm}

If ${\sum\nolimits_{k = 1}^K {p_k^ \star (0 )} > {P_{\max }}} $, we need to find a $\lambda$ to satisfy Equation (\ref{sdefrgef}).
For the case of ${p_k^\star}(\lambda)>0$, by taking the first-order derivative of  ${p_k^\star}(\lambda)$ w.r.t. $\lambda$, we have $\frac{{\partial {p_k^\star}(\lambda )}}{{\partial \lambda }} < 0$.

\subsection{The Solution of $\text{(P2-2)}$}\label{fjhjtkeho}

 In this subsection, our aim is to solve Problem $\text{(P2-2)}$ by optimizing the number of bandwidth units with given power allocation. For simplicity, we first define ${{\tilde N}_0} = {B_0}T,\tilde g_k^d = {p_k}g_k^d,\tilde g_k^e = {p_k}{g^e},$ ${{\tilde L}_k^d} = \sqrt {{{\tilde N}_0}} \frac{{{Q^{ - 1}}({\epsilon_k})}}{{\ln 2}}$, and ${{\tilde L}_k^e} = \sqrt {{{\tilde N}_0}} \frac{{{Q^{ - 1}}({\sigma _k})}}{{\ln 2}}$. Since $g_k^d>g^e$, we have $\tilde g_k^d>\tilde g_k^e$. Then,  $R_k(n_k)$ can be rewritten as
 \begin{equation}\label{defreg}
   {R_k}({n_k}) = \underbrace {{{\tilde N}_0}{n_k}{{\log }_2}\left( {1 + \frac{{\tilde g_k^d}}{{{n_k}}}} \right) - {{\tilde N}_0}{n_k}{{\log }_2}\left( {1 + \frac{{\tilde g_k^e}}{{{n_k}}}} \right)}_{F_k({n_k})} - \underbrace {\left( {\sqrt {z_k^d({n_k})} {{\tilde L}_k^d} + \sqrt {z_k^e({n_k})} {{\tilde L}_k^e}} \right)}_{G_k({n_k})}
 \end{equation}
 where $z_k^x({n_k}) = {n_k} - \frac{{n_k^3}}{{{{\left( {{n_k} + \tilde g_k^x} \right)}^2}}}$, $x\in \{d,e\}$.

 Before solving Problem $\text{(P2-2)}$, we first analyze the concave-convex property of  ${R_k}({n_k})$. In particular, the following lemma shows that ${F_k({n_k})}$ and ${G_k({n_k})}$ are concave functions w.r.t. $n_k$.

\emph{\bf{Lemma 2}}:  ${F_k({n_k})}$ and ${G_k({n_k})}$ are concave w.r.t. $n_k$, and thus $R_k(n_k)$ is the difference of two concave functions ${F_k({n_k})}$ and ${G_k({n_k})}$.

\emph{Proof}: Please refer to Appendix \ref{lemma2}.   \hfill\rule{2.7mm}{2.7mm}

Then, similar to the optimization of power allocation, we adopt the SCA method to solve  Problem $\text{(P2-2)}$. By denoting the solution $\bm{n}$ in the $(j-1)$-th iteration as $\bm{n}^{(j-1)}$ and using Lemma 2 and Jensen's inequality, we have
\begin{equation}\label{sdefwgt}
 {G_k}({n_k}) \le {G_k}(n_k^{(j - 1)}) + {\alpha _k}(n_k^{(j - 1)})\left( {{n_k} - n_k^{(j - 1)}} \right),
\end{equation}
where ${\alpha _k}(n_k^{(j - 1)})$ is the first-order derivative of ${G_k}({n_k})$ w.r.t. $n_k$ at $n_k=n_k^{(j - 1)}$ and is  given by
\begin{equation}\label{efggtrh}
  {\alpha _k}(n_k^{(j - 1)}) = \frac{{{{\tilde L}_k^d}\left( {3{{\left( {\tilde g_k^d} \right)}^2}n_k^{(j - 1)} + {{\left( {\tilde g_k^d} \right)}^3}} \right)}}{{2\sqrt {z_k^d(n_k^{(j - 1)})} {{\left( {n_k^{(j - 1)} + \tilde g_k^d} \right)}^3}}} + \frac{{{{\tilde L}_k^e}\left( {3{{\left( {\tilde g_k^e} \right)}^2}n_k^{(j - 1)} + {{\left( {\tilde g_k^e} \right)}^3}} \right)}}{{2\sqrt {z_k^e(n_k^{(j - 1)})} {{\left( {n_k^{(j - 1)} + \tilde g_k^e} \right)}^3}}}.
\end{equation}

By  replacing ${\alpha _k}(n_k^{(j - 1)})$ with the RHS of (\ref{efggtrh}), the subproblem to be solved in the $j$th iteration is given by
\begin{subequations}
\begin{align}
({\bf{P2-2-a}}): \;\;\max_{{\bm{n}}}\;\;\;& \sum\nolimits_{k = 1}^K {\left( {\omega _kF_k({n_k}) - \omega _k{\alpha _k}(n_k^{(j - 1)}){n_k}} \right)}  \\
\text{s.t.}\;\;\;\;\;\;&(\ref{blocklength}), (\ref{blockldfh}),
\end{align}
\end{subequations}
where the constant terms are omitted in the objective function.

Note that the objective function of Problem $({\bf{P2-2-a}})$ is a concave function of $n_k$ and the constraints are affine functions. Hence, Problem $({\bf{P2-2-a}})$ is a convex optimization problem. In the following, we provide a low-complexity algorithm to obtain the globally optimal solution by using the Lagrangian dual decomposition method \cite{boyd2004convex}. Since Problem $({\bf{P2-2-a}})$ is a convex optimization problem and the slater's condition is satisfied \footnote{There exist strictly feasible $\bm{n}$ such as $\sum\nolimits_{k=1}^Kn_k< n_{\rm{max}}$.}, the dual gap is zero and the original problem can be solved by solving its dual problem. Specifically, we introduce the non-negative   Lagrange multiplier $\mu\ge 0$ corresponding to the constraint of the total number of bandwidth units, and the partial Lagrange function of Problem $({\bf{P2-2-a}})$ is given by
\begin{equation}
\mathcal{L} ({\bm{n}},\mu)=\sum\nolimits_{k = 1}^K {\left( {\omega _kF_k({n_k}) - \omega _k{\alpha _k}(n_k^{(j - 1)}){n_k}} \right)} - \mu \left( {\sum\nolimits_{k = 1}^K {{n_k}}  - {n_{\max }}} \right).
\end{equation}
The dual function can be obtained by solving the following optimization problem:
\begin{equation}\label{sdwefgrt}
Y(\mu ) \buildrel \Delta \over = \mathop {\max }\limits_{{n_k} \ge 0,\forall k} \sum\nolimits_{k = 1}^K {{\mathcal{L}_k}({n_k},\mu )}+\mu{n_{\max }},
\end{equation}
where ${\mathcal{L}_k}({n_k},\mu )$ is given by
\begin{equation}\label{sdfrgth}
  {\mathcal{L}_k}({n_k},\mu )= {\omega _kF_k({n_k}) - \omega _k{\alpha _k}(n_k^{(j - 1)}){n_k}}  - \mu {n_k}.
\end{equation}
Then, the dual problem is given by
 \begin{equation}\label{sdefrg}
  \mathop {\min }\limits_{\mu \ge 0} Y(\mu).
 \end{equation}

To solve the dual problem (\ref{sdefrg}), we need to first obtain the expression of dual function $Y(\mu)$, which needs to solve
 Problem (\ref{sdwefgrt}) with given $\mu$. For given $\mu$, Problem (\ref{sdwefgrt}) can be decoupled into $K$ independent optimization problems, and the optimization problem associated with the $k$th device is given by
 \begin{equation}\label{dwreg}
  \mathop {\max }\limits_{{n_k} \ge 0} \;{\mathcal{L}_k}({n_k},\mu )= {\omega _kF_k({n_k}) - \omega _k{\alpha _k}(n_k^{(j - 1)}){n_k}}  - \mu {n_k}.
 \end{equation}
The first-order derivative of ${\mathcal{L}_k}({n_k},\mu )$ w.r.t. $n_k$ is derived as
\begin{equation}\label{dwfgtr}
 \frac{{\partial \mathcal{L}_k({n_k},\mu )}}{{\partial {n_k}}}\!=\omega _k\tilde{N}_0\log_2\left(\frac{n_k+\tilde{g}_k^d}{n_k+\tilde{g}_k^e}\right)+\! \frac{{{\omega _k{\tilde N}_0}\left( {\tilde g_k^e \!-\! \tilde g_k^d} \right){n_k}}}{{\ln2\left( {{n_k} \!+ \!\tilde g_k^d} \right)\left( {{n_k}\! +\! \tilde g_k^e} \right)}}\!-\!\omega _k{\alpha _k}\left(n_k^{(j - 1)}\right)
- \mu.
\end{equation}
Since ${\mathcal{L}_k}({n_k},\mu )$ is a concave function, the optimal solution of Problem (\ref{dwreg}) can be derived as follows:

$\bullet$ If ${\left. {\frac{{\partial {{\cal L}_k}({n_k},\mu )}}{{\partial {n_k}}}} \right|_{{n_k} = 0}} \le 0$, the optimal ${n_k}$ for given $\mu$ is given by ${n_k^\star}(\mu)=0$;

$\bullet$ If ${\left. {\frac{{\partial {{\cal L}_k}({n_k},\mu )}}{{\partial {n_k}}}} \right|_{{n_k} = 0}} > 0$, the optimal ${n_k}$ should satisfy the equation
  $\frac{{\partial {{\cal L}_k}({n_k},\mu )}}{{\partial {n_k}}} = 0$, and its root is denoted as $n_k^\star(\mu)$. As $\mathcal{L}_k(n_k,\mu)$ is a concave function,  $\frac{{\partial \mathcal{L}_k({n_k},\mu )}}{{\partial {n_k}}}$ is a decreasing function w.r.t. $n_k$. Thus, $n_k^*(\mu)$ can be obtained by the bisection search method when $\mu$ is given.

Next, we turn to solve the dual problem by finding the optimal $\mu$. The optimal value of $\mu$ should satisfy the complementary slackness condition for the constraint (\ref{blocklength}):
\begin{equation}\label{dfgetr}
  \mu \left( {\sum\nolimits_{k = 1}^K {{n_k^\star}(\mu )}  - {n_{\max }}} \right) = 0.
\end{equation}

From (\ref{dfgetr}), if $\sum\nolimits_{k = 1}^K {{n_k^\star}(0 )}  \le {n_{\max }}$ holds, the optimal solution is given by ${n_k}(0 ),\forall k$; Otherwise, the optimal $\mu$ should satisfy $W(\mu)=\sum\nolimits_{k = 1}^K {{n_k^\star}(\mu )} = {n_{\max }}.$ In contrast to the power allocation solution in (\ref{defrfe}), the bandwidth unit allocation $n_k(\mu)$ cannot be expressed in an explicit function of $\mu$. As a result, the monotonicity of $n_k(\mu)$ w.r.t. $\mu$  cannot be proved by checking the sign of the first-order derivative. To deal with this issue, we have the following lemma:

\emph{\bf{Lemma 3}}:  $W(\mu)$ is a monotonically decreasing function w.r.t. $\mu$.

\emph{Proof}: Please refer to Appendix \ref{lemma3}.   \hfill\rule{2.7mm}{2.7mm}

Based on Lemma 3, the optimal $\mu$ can be obtained by using the bisection search method.


\subsection{Algorithm Analysis}
\subsubsection{Algorithm Description}
Based on the above analysis, we summarize the proposed BCD algorithm in Algorithm \ref{hgirjoj}, where $R\left( {{{\bm{n}}},{{\bm{p}}}} \right)$ is the weighted throughput defined as
$R\left( {{{\bm{n}}},{{\bm{p}}}} \right) = \sum\nolimits_{k = 1}^K {\omega _k{R_k}\left( {{{\bm{n}}},{{\bm{p}}}} \right)}$. This algorithm is a two-layer iterative algorithm, where the inner layer is the iteration of the SCA algorithm to solve Problem  $({\bf{P2-1-a}})$ and Problem $({\bf{P2-2-a}})$, and the outer layer is the BCD algorithm to solve Problem  $\textbf{(P2)}$.
In Line 7 of Algorithm \ref{hgirjoj}, ${\bm{p}}^\star$ denotes the optimal solution obtained by the inner layer to solve Problem  $({\bf{P2-1-a}})$, and ${\bm{n}}^\star$ in Line 12 corresponds to the inner layer to solve Problem $({\bf{P2-2-a}})$.
\begin{algorithm}
\setstretch{0.5}
  \caption{BCD Algorithm for Solving Total Throughput Maximization}\label{hgirjoj}
   \textbf{Initialize}  ${\bm{n}}={\bm{n}}^{(0)}$, ${\bm{p}}={\bm{p}}^{(0)}$, accuracy $\varepsilon$, the iteration number $t=1$ and calculate ${R\left( {{{\bm{n}}^{(0)}},{{\bm{p}}^{(0)}}} \right)}$;

  \Repeat { ${{\left| {R\left( {{{\bm{n}}^{(t)}},{{\bm{p}}^{(t)}}} \right) - R\left( {{{\bm{n}}^{(t - 1)}},{{\bm{p}}^{(t - 1)}}} \right)} \right|} \mathord{\left/
 {\vphantom {{\left| {R\left( {{{\bm{n}}^{(t)}},{{\bm{p}}^{(t)}}} \right) - R\left( {{{\bm{n}}^{(t - 1)}},{{\bm{p}}^{(t - 1)}}} \right)} \right|} {R\left( {{{\bm{n}}^{(t - 1)}},{{\bm{p}}^{(t - 1)}}} \right) \le }}} \right.
 \kern-\nulldelimiterspace} {R\left( {{{\bm{n}}^{(t - 1)}},{{\bm{p}}^{(t - 1)}}} \right) \le }}\varepsilon $ }
  {Set ${\bm{n}}={\bm{n}}^{(t-1)}$, $i=1$;

  \Repeat{$\bm{p}$ converges}
  {Given ${\bm{p}}^{(i-1)}$, calculate ${\bm{p}}^{(i)}$ by solving Problem $({\bf{P2-1-a}})$, and $i \leftarrow i + 1$;}
  {Update ${\bm{p}}^{(t)}={\bm{p}}^\star$;}

  {Set ${\bm{p}}={\bm{p}}^{(t)}$, $j=1$;}

  \Repeat{$\bm{n}$ converges}
  {Given ${\bm{n}}^{(j-1)}$, calculate ${\bm{n}}^{(j)}$ by solving Problem $({\bf{P2-2-a}})$, and $j \leftarrow j + 1$;}

  {Update ${\bm{n}}^{(t)}={\bm{n}}^\star$ and set $t \leftarrow t + 1$;}
  }

\end{algorithm}

\subsubsection{Convergence Analysis}
In the $t$th outer iteration, the SCA algorithm is adopted to solve Problem $({\bf{P2-1-a}})$ to find the power allocation solution.
Based on the property of SCA algorithm \cite{cunhua2019}, the SCA algorithm is guaranteed to converge. Then, we have $R\left( {{{\bm{n}}^{(t-1)}},{{\bm{p}}^{(t)}}} \right) \ge R\left( {{{\bm{n}}^{(t - 1)}},{{\bm{p}}^{(t - 1)}}} \right)$. Afterwards, the SCA
algorithm is used to find the channel bandwidth unit solution. We then have $R\left( {{{\bm{n}}^{(t)}},{{\bm{p}}^{(t)}}} \right) \ge R\left( {{{\bm{n}}^{(t - 1)}},{{\bm{p}}^{(t )}}} \right)$. Hence, we have $R\left( {{{\bm{n}}^{(t)}},{{\bm{p}}^{(t)}}} \right) \ge R\left( {{{\bm{n}}^{(t - 1)}},{{\bm{p}}^{(t-1 )}}} \right)$, which shows the solutions obtained by the BCD algorithm are monotonically increasing. In addition, due to the power and total bandwidth limits, there exists an upper bound on the total throughput. Hence, the BCD algorithm is guaranteed to converge.

\subsubsection{Complexity Analysis}
In this part, we analyze the complexity of the BCD algorithm. Note that the main complexity in each outer layer iteration lies in the SCA algorithms to solve Problem $({\bf{P2-1-a}})$ and Problem $({\bf{P2-2-a}})$.  For each inner layer of the SCA algorithm to solve $({\bf{P2-1-a}})$, the bisection search method is adopted to find $\lambda$, and its complexity is ${\cal O}\left( {K{{\log }_2}\left( {\frac{1}{\varepsilon }} \right)} \right)$, where $\varepsilon $ is the accuracy. Denote $I_{\rm{in}}$ as the total number of iterations  required for the convergence of the SCA algorithm. The total complexity to solve $({\bf{P2-1-a}})$ in each outer layer is given by
${\cal O}\left(I_{\rm{in}} {K{{\log }_2}\left( {\frac{1}{\varepsilon }} \right)} \right)$. By using similar analysis, the total complexity to solve $({\bf{P2-2-a}})$ in each outer layer is given by ${\cal O}\left(J_{\rm{in}} {K{{\log }_2}\left( {\frac{1}{\varepsilon }} \right)} \right)$, where $J_{\rm{in}}$ is the total number of iterations required for the convergence of the SCA algorithm. Denote the total number of iterations for the BCD algorithm to converge as $N_{\rm{BCD}}$. The total complexity of the BCD algorithm is given by ${\cal O}\left({N_{{\rm{BCD}}}}\left( {{I_{{\rm{in}}}} + {J_{{\rm{in}}}}} \right)K{\log _2}\left( {\frac{1}{\varepsilon }} \right) \right)$. Hence, the BCD algorithm can converge to the locally optimal solution in polynomial
time computational complexity.

\subsection{Integer Conversion for $\bm{n}$}
In general, the solution of $\bm{n}$ obtained from the BCD algorithm is positive continuous values, which may violate the integer constraints. In this part, we provide a greedy search method to convert the continuous solution into integer ones. Specifically, denote the solution of $\bm{n}$ obtained by the BCD algorithm as $\bar{\bm{n}}=\{\bar n_1, \cdots, \bar n_K\}$. The integer conversion problem is a combinatorial optimization problem, and it requires exponential time complexity to find the globally optimal solution. In the following, we propose a low-complexity algorithm based on the greedy search method to find a suboptimal solution. Firstly, we set the initial value of the solution as $n_k^\star = \left\lfloor {{{\bar n}_k}} \right\rfloor, \forall k$, where $\left\lfloor \cdot \right\rfloor $ denotes the flooring operation. Then, there are $N_{\rm{Rem}}=\sum\nolimits_{k = 1}^K {{{\bar n}_k}}  - \sum\nolimits_{k = 1}^K {{n_k^\star}}$ bandwidth units that are not allocated. The remaining task is to allocate these bandwidth units to the devices. The main idea of the greedy search method is that each time we allocate one bandwidth unit to the device with the highest increment of the total throughput. Denote $ {\bm{n}}^\star=\{ n_1^\star, \cdots,  n_K^\star\}$ and $ {\bm{{\tilde n}}}_k=\{ n_1^\star, \cdots, n_k^\star+1, \cdots,   n_K^\star\}$. For each given $\bm{n}$, we adopt the SCA algorithm to solve  Problem $({\bf{P2-1}})$, and denote the optimal value of the total throughput as $R\left( {\bm{n}} \right)$. Then, the device index to be allocated one bandwidth unit is given by ${k^*} = \arg {\max _{k \in {\cal K}}}\left\{ {R\left( {\bm{{\tilde n}}}_k \right) - R\left( {\bm{n}}^\star \right)} \right\}$, where ${\cal K}$ denotes the set of all devices. For the $k^*$ device, update $n_k^\star=n_k^\star+1$. Repeat the above procedure until all the remaining bandwidth units are allocated, and the power allocation is updated accordingly based on the final integer solution.

\section{Total Transmit Power Minimization}\label{ttpmin}

In this section, each device is assumed to have its minimum throughput requirement, and our goal is to minimize the TTP by jointly optimizing the bandwidth unit and power allocation. We first provide the problem formulation and then propose one efficient algorithm to solve the problem.

\subsection{Problem Formulation}

In some application scenarios where the power consumption of the AP is of great concern, the design paradigm should be shifted to the energy efficient design by minimizing the power consumption. Specifically, we aim to jointly optimize the bandwidth unit and power allocation to  minimize the TTP while guaranteeing each device's minimum throughput requirement and the budget of  the total available bandwidth units. Mathematically, this optimization problem is formulated as follows:
\begin{subequations}\label{vojgtjjie}
\begin{align}
({\bf{P3}}):\;\;\min_{{\bm{p}},{\bm{n}}}\;\;\;&  \sum\nolimits_{k=1}^{K} p_k  \\
\text{s.t.}\;\;\;\;\;\; &R_k\ge D_k^{\min}, \forall k,\label{dkncjo}\\
&(\ref{blocklength}), (\ref{blocklength-cons}), (\ref{nonegp}),
\end{align}
\end{subequations}
where $D_k^{\min}$ is the minimum throughput requirement of the $k$th device. In the following, we always assume that the problem is feasible.

Problem $({\bf{P3}})$ can be readily known as a mixed-integer programming problem  due to the integer constraint on the number of bandwidth units, which is NP-hard to solve. We notice that the objective function of Problem $({\bf{P3}})$ is not related to the number of bandwidth units and only depends on the power allocation. Hence, the BCD algorithm that alternatively optimizes the bandwidth unit and power allocation is not applicable. In the following, we assume that the problem is feasible and we propose one low-complexity algorithm to solve this problem.

\subsection{Approximation Method}

The complicated expression of data rate $R_k$ makes Problem (\ref{vojgtjjie}) difficult to solve. To make it tractable, we approximate $V_k^x$ as one, i.e., $V_k^x\approx 1$, where $x\in\{d,e\}$. The approximation is very accurate when SNR rate $\gamma_k^x$ is very high, $\gamma_k^x\gg1$. This approximation has been widely adopted in the current literature \cite{changyang2018,changjian}. Define
$\tilde h_k^d = {{T{{\left| {{h_k}} \right|}^2}} \mathord{\left/
 {\vphantom {{T{{\left| {{h_k}} \right|}^2}} {\sigma _{d,k}^2}}} \right.
 \kern-\nulldelimiterspace} {\sigma _{d,k}^2}}$ and  $\tilde h^e = T{{\left\| {{{\bf{h}}_e}} \right\|_2^2} \mathord{\left/
 {\vphantom {{\left\| {{{\bf{h}}_e}} \right\|_2^2} {\sigma _e^2}}} \right.
 \kern-\nulldelimiterspace} {\sigma _e^2}}$ with $\tilde h_k^d>\tilde h^e$, the achievable data rate can be approximated as
\begin{equation}\label{hdjhgt}
  R_k \approx {{\tilde R}_k}\!=\!N_k\left(  {  \log_2\left(1\!+\!\frac{{{p_k}{{\tilde h}_k^d}}}{{{N_k}}}\right)\!-\!\log_2\left(1+\frac{{{p_k}{{\tilde h}^e}}}{{{N_k}}}\right)\!-\!\sqrt {\frac{{1}}{{N_k}}} \frac{{{Q^{ - 1}}({\epsilon_k})}}{{\ln 2}} \!\! -\!\! \sqrt {\frac{{1}}{{N_k}}} \frac{{{Q^{ - 1}}({\delta _k})}}{{\ln 2}}}  \right).
\end{equation}
Since $V_k^x<1$, ${{\tilde R}_k}$ is actually a lower bound of the original data rate $R_k$. Hence, if ${\tilde R}_k\ge D_k^{\min}$, then $R_k\ge D_k^{\min}$ always holds.

By substituting (\ref{hdjhgt}) into Problem $({\bf{P3}})$, we can now optimize the channel blocklength allocation ${\bm{N}}({\bm{N}}=\{N_1,\cdots,N_K\})$ and the power allocation ${\bm{p}}$, which is formulated  as
\begin{subequations}\label{vaxasjjie}
\begin{align}
({\bf{P4}}):\;\;\min_{{\bm{p}},{\bm{N}}}\;\;\;&  \sum\nolimits_{k=1}^{K} p_k \\
\text{s.t.}\;\;\;\;\;\; &{\tilde R}_k\ge D_k^{\min}, \forall k,\label{dknsacdsjo}\\
&\sum\nolimits_{k = 1}^K {{N_k}}  \le {W_c}T,\label{joijijie}\\
&p_k\ge 0, \forall k \label{joierj}\\
&{N_k} \in \{ {B_0}T,2{B_0}T, \cdots ,{n_{\max }}BT\}, \forall k, \label{wfoho}
\end{align}
\end{subequations}
where $W_c$ is the coherence channel bandwidth. Due to the  discrete constraint on $N_k$, Problem $({\bf{P4}})$ is difficult to solve. To solve this problem, we first remove this constraint and relax it to continuous values, which is given by
\begin{subequations}\label{vaxasjie}
\begin{align}
({\bf{P4-a}}):\;\;\min_{{\bm{p}},{\bm{N}}}\;\;\;&  \sum\nolimits_{k=1}^{K} p_k \\
\text{s.t.}\;\;\;\;\;\; & (\ref{dknsacdsjo}), (\ref{joijijie}), (\ref{joierj}),\\
&N_k\ge 0. \label{fcjegtog}
\end{align}
\end{subequations}
When Problem ({\bf{P4-a}}) is solved, we convert the continuous $N_k$s into discrete values.

 We first solve Problem ({\bf{P4-a}}). Obviously, for any given channel blocklength allocation $N_k$, ${\tilde R}_k$ is a monotonically increasing function of $p_k$. Hence, inequality (\ref{dknsacdsjo}) holds with equality at the optimal point. Then, the power allocation can be expressed as a function of $N_k$:
\begin{equation}\label{yhufyk}
 {p_k}({N_k}) =  - \frac{{{N_k}}}{{{{\tilde h}^e}}} + \frac{{c_k{N_k}}}{{d_k - {e^{\frac{a_k}{{{N_k}}} + \frac{b_k}{{\sqrt {{N_k}} }}}}}},
\end{equation}
where $a_k = D_k^{\min }\ln 2,b_k = {Q^{ - 1}}({\epsilon_k}) + {Q^{ - 1}}({\delta _k})$, $c_k = {{\left( {\tilde h_k^d - {{\tilde h}^e}} \right)} \mathord{\left/
 {\vphantom {{\left( {\tilde h_k^d - {{\tilde h}^e}} \right)} {{{\left( {{{\tilde h}^e}} \right)}^2}}}} \right.
 \kern-\nulldelimiterspace} {{{\left( {{{\tilde h}^e}} \right)}^2}}}$, and $d_k = {{\tilde h_k^d} \mathord{\left/
 {\vphantom {{\tilde h_k^d} {{{\tilde h}^e}}}} \right.
 \kern-\nulldelimiterspace} {{{\tilde h}^e}}}$.  To guarantee that $N_k$ is  positive, by using ${p_k}({N_k})>0$, we can obtain the lower bound of $N_k$ as
 \begin{equation}\label{hmkfjy}
   {N_k} \ge {\left( {\frac{{{b_k} + \sqrt {b_k^2 + 4{a_k}\ln {d_k}} }}{{2\ln {d_k}}}} \right)^2} \buildrel \Delta \over = N_k^{{\rm{lb}}}.
 \end{equation}

 By substituting (\ref{yhufyk}) into Problem ({\bf{P4-a}}) and considering the lower bound of $N_k$, we have
 \begin{subequations}\label{vaaxswajie}
\begin{align}
({\bf{P4-b}}):\;\;\min_{{\bm{N}}}\;\;\;&  \sum\nolimits_{k=1}^{K} p_k(N_k)  \\
\text{s.t.}\;\;\;\;\;\; & (\ref{joijijie}), (\ref{hmkfjy}).
\end{align}
\end{subequations}

 Then, in the following theorem, we provide a sufficient condition when ${p_k}({N_k})$ is a convex function.

\emph{\textbf{Theorem 2}}: By defining ${\rho _k} =  - \frac{{12{a_k} + b_k^2}}{3}$ and ${\kappa _k} = -\frac{{2b_k^4 + 36{a_k}b_k^2 + 108a_k^2}}{{27{b_k}}}$, the function  ${p_k}({N_k})$ is a monotonically decreasing and convex function w.r.t. $N_k$ when the following condition holds:
 \vspace{-0.2cm}
 \begin{equation}\label{fdvtgxdtr}
 \vspace{-0.2cm}
  \sqrt {{N_k}}  \le   2\sqrt { - \frac{{{\rho _k}}}{3}} \cosh \left( {\frac{1}{3}{\rm{arcosh}}\left( {\frac{{   3{\kappa _k}}}{{2{\rho _k}}}\sqrt {\frac{{ - 3}}{{{\rho _k}}}} } \right)} \right)+\frac{{{b_k}}}{3}.
 \end{equation}
\emph{Proof}: \upshape Please see Appendix \ref{theorem2}. \hfill\rule{2.7mm}{2.7mm}

Fortunately, the RHS of (\ref{fdvtgxdtr}) only depends on the long-term system parameters such as $D_k^{\min }$, $\epsilon_k$, and $\delta _k$, which is not related to the relative channel gains. For a typical URLLC communication system, the number of transmission bits for each device is around 100 bits (i.e., $D_k^{\min }=100$), the decoding error probability $\epsilon_k$ is about $10^{-9}$, the information leakage  $\delta_k$  is roughly $10^{-2}$. Then, the value of the RHS of (\ref{fdvtgxdtr}) can be calculated as  23.9. Hence, when $N_k\le 572$, the inequality in (\ref{fdvtgxdtr}) holds. For a typical system, the channel coherence bandwidth is around 0.5 MHz, and the transmission delay requirement  is 1 ms. Hence, the total number of channel uses is 500, which should be allocated among all devices. Then, the number of channel uses allocated to each device is much smaller than the value of 500. As a result, for practical communication systems, the inequality in (\ref{fdvtgxdtr}) holds and thus ${p_k}({N_k})$ is a  convex function.

Since constraints (\ref{joijijie}) and  (\ref{hmkfjy}) are affine functions, Problem $({\bf{P4-b}})$ is a convex problem, which can be solved by using Lagrangian dual decomposition method. We first introduce a positive Lagrange multiplier $\varsigma$ associated with constraint (\ref{joijijie}), the partial Lagrangian function of Problem $({\bf{P4-b}})$ is given by
\begin{equation}
\mathcal{L}({\bm{N}},\varsigma)=  {\sum\nolimits_{k = 1}^K {{p_k}} ({N_k})}   + \varsigma \left( {\sum\nolimits_{k = 1}^K {{N_k}}  - {W_c}T} \right).
\end{equation}
We first need to obtain the optimal ${\bm{N}}$  by minimizing $\mathcal{L}({\bm{N}},\varsigma)$ over ${\bm{N}}$ for a given $\varsigma$:
\begin{equation}\label{ghyte}
  \min_{{\bm{N}}}\mathcal{L}({\bm{N}},\varsigma).
\end{equation}
 We denote the optimal $N_k$ for given $\varsigma$ as $N_k^\star(\varsigma)$.
For given $\varsigma$, $\mathcal{L}({\bm{N}},\varsigma)$ is a convex function, and thus  the optimal $N_k^\star(\varsigma)$ can be obtained as follows:
\begin{itemize}
  \item If $\frac{\partial \mathcal{L}({\bm{N}},\varsigma)}{\partial {N}_k}|_{N_k=N_k^{\rm{lb}}}\ge 0$, the optimal $N_k$ is given by $N_k^\star(\varsigma)=N_k^{\rm{lb}}$;
  \item If $\frac{\partial \mathcal{L}({\bm{N}},\varsigma)}{\partial {N}_k}|_{N_k=N_k^{\rm{lb}}}< 0$, $N_k^\star(\varsigma)$ is the solution to the equation $\frac{\partial \mathcal{L}({\bm{N}},\varsigma)}{\partial {N}_k}=0$, which can be obtained by bisection search method.
\end{itemize}

Once obtaining the optimal $N_k^\star(\varsigma)$, we can obtain the sum of all channel uses defined as
\begin{equation}
F(\varsigma)\triangleq {\sum\nolimits_{k = 1}^K {N_k^ \star (\varsigma )} }.
\end{equation}
We need to solve the following equation to find the optimal dual variable $\varsigma$:
\begin{equation}\label{kugyj}
  \varsigma \left( {F(\varsigma ) - {W_c}T} \right)=0.
\end{equation}
If $F(0)\le {W_c}T$, then the optimal $\varsigma$ is equal to zero. Otherwise, we need to solve the equation $F(\varsigma )={W_c}T$.
By using a similar method as in Lemma 3, we can prove that  $F(\varsigma)$ is a monotonically decreasing function of $\varsigma$. Hence, the bisection search method can be used to find the solution of equation $F(\varsigma )={W_c}T$.

Denote the solution obtained from Problem $({\bf{P4-b}})$ as  $\bm{\bar N}=\{\bar N_1,\cdots,\bar N_K\}$. Obviously, the solution $\bm{\bar N}$ obtained by using the above the  Lagrangian dual decomposition method do not satisfy the discrete constraint in (\ref{wfoho}). Hence, we need to transfer $\bm{\bar N}$ to satisfy its discrete constraint. As mentioned before, this kind of problem is a combinatorial optimization problem, which is NP-hard to solve. We again adopt the greedy search method to solve this problem.
Denote the solution of $\bm{N}$ that satisfies the discrete constraint as $\bm{N^\star}=\{N_1^\star,\cdots,N_K^\star\}$. Specifically, we first initialize the solution of $\bm{N^\star}$ as ${N_k^\star} = \left\lfloor {\frac{{{{\bar N}_k}}}{{{B_0}T}}} \right\rfloor \cdot {B_0}T, \forall k $. There are other channel uses that have not allocated, the number of which is given by $\left( {{n_{\max }} - \sum\limits_{k = 1}^K {\left\lfloor {\frac{{{{\bar N}_k}}}{{{B_0}T}}} \right\rfloor } } \right) \cdot {B_0}T$.  As proved in Theorem 2, ${p_k}({N_k})$ is a monotonically decreasing function of $N_k$. Hence, we can assign the unallocated channel uses to additionally reduce the power consumption. We allocate one channel use to the device with the largest decrement of ${p_k}({N_k})$, i.e., $k^\star=\arg \mathop {\max }\limits_{k \in {\cal K}} \{p_k(N_k^\star)-p_k(N_k^\star+B_0T)\}$. For the $k^\star$th device, we allocate one bandwidth unit to it and update $N_k^\star=N_k^\star+B_0T$. Repeat this procedure until $\sum\nolimits_{k = 1}^K {N_k^\star}  = {W_c}T$.

\section{Simulation Results}\label{simul}
In this section, we provide simulation results to evaluate the performance of our proposed algorithms. Unless specified otherwise, the adopted simulation parameters are given as follows: bandwidth of channel unit of $B_0=1$ KHz, noise power  spectrum density of -173 dBm/Hz, number of devices of $K=4$,    $\epsilon_k=10^{-9}, \forall k$, $\delta_k=10^{-2}, \forall k$, time duration of $T=1$ ms, channel coherence bandwidth of $W_c=0.5$ MHz. The channel path loss is modeled as $PL=35.3+37.6{\rm{log}}_{10}l$ (dB) \cite{access2010further}, where $l$ (m) is the distance between the devices/eavesdropper and the AP. The distance between the eavesdropper and the AP is set as $l_e=180$ (m).

\subsection{Weighted Sum Throughput Maximization}
In this subsection, we provide simulation results to evaluate the performance of the BCD algorithm in Algorithm \ref{hgirjoj} for the WST maximization problem. The distances between the AP and the devices are assumed to be randomly generated within $100$ m $\sim $ $120$ m, and the following results are obtained by averaging over 200 device location generations.

\begin{figure}
\vspace{-0.3cm}
\begin{minipage}[t]{0.5\linewidth}
\setcaptionwidth{\textwidth}
\captionsetup{font=small}
\centering
\includegraphics[width=\linewidth]{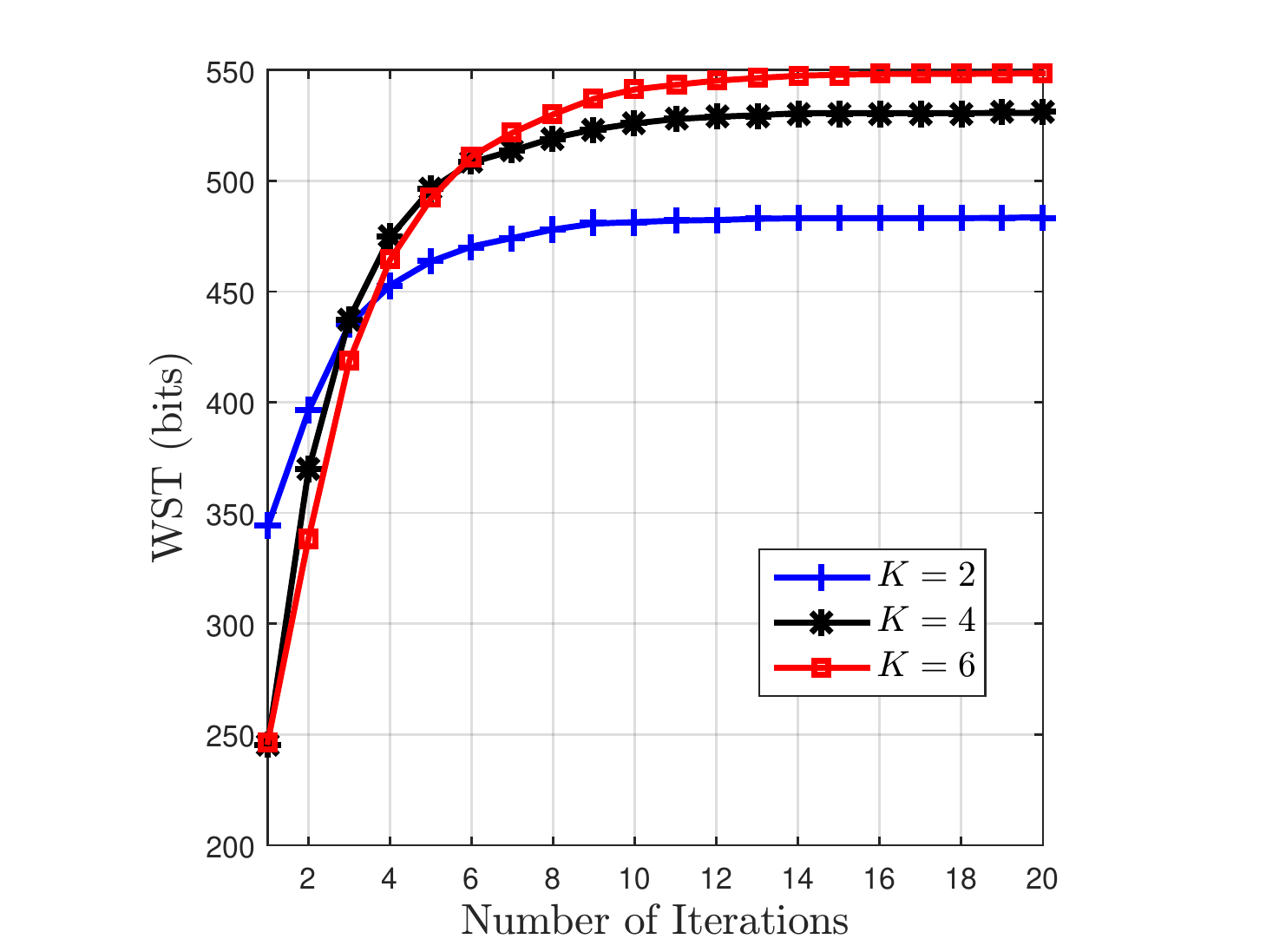}
\vspace{-0.3cm}
\caption{Convergence behavior of the BCD algorithm.}
\label{conver}
\vspace{-0.4cm}
\end{minipage}
\begin{minipage}[t]{0.5\linewidth}
\setcaptionwidth{\textwidth}
\captionsetup{font=small}
\centering
\includegraphics[width=\linewidth]{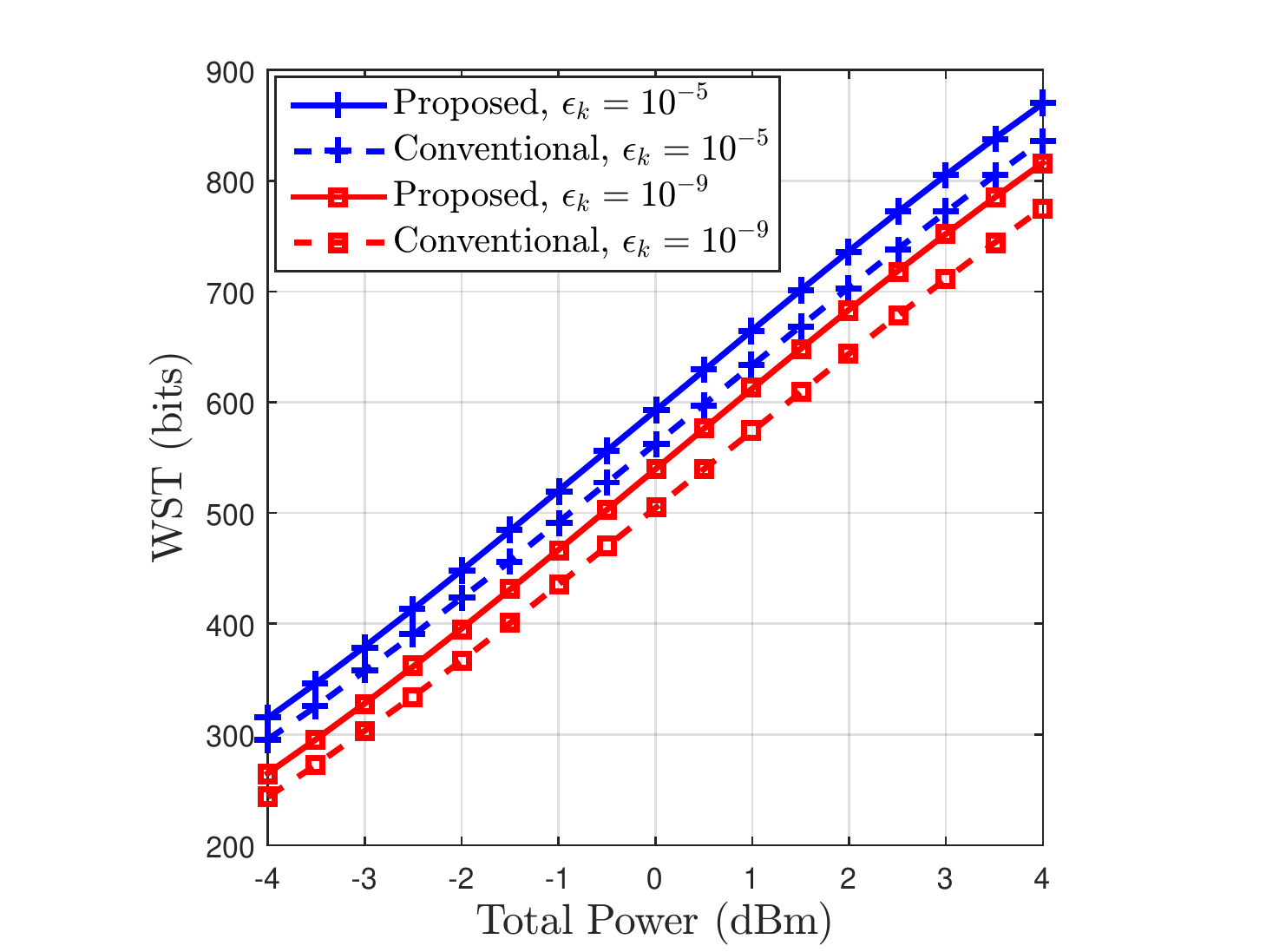}
\vspace{-0.3cm}
\caption{WST versus the total power limit.}
\vspace{-0.4cm}
\label{throuvspower}
\end{minipage}
\end{figure}

In Fig.~\ref{conver}, we illustrate the convergence behavior of the BCD algorithm for various number of devices. It is observed from this figure that the BCD algorithm converges rapidly for all considered values of $K$, and roughly ten iterations are sufficient for the convergence of the BCD algorithm. Fig.~\ref{conver} also shows that larger number of devices leads to slower convergence speed. The reason is that larger number of devices corresponds to more optimization variables to be optimized and require more iterations.

Next, we compare the performance of the proposed BCD algorithm with the conventional long packet transmission, where the penalty terms in (\ref{reteexpression}) are not considered and the throughput of the $k$th device is given by
\begin{equation}\label{dwsfr}
 R_k= {n_k}{B_0}T \left( \log_2(1+\gamma^d_k)-\log_2(1+\gamma^e_k)\right).
\end{equation}
The BCD algorithm can be directly applied by setting some parameters to zero. This algorithm is labeled as `Conventional'.

Fig.~\ref{throuvspower} shows the WST versus the total power limit for various decoding error probabilities at the devices. As expected, the WST of each algorithm increases with the increase of the maximum available transmit power as higher transmit power will bring higher value of SNR. The proposed BCD algorithm is observed to outperform the conventional long packet transmission scheme, and the performance gap increases with increasing the transmit power limit. This may be due to the fact that larger transmit power corresponds to a higher value of SNR, and thus $V_k^x$ will approach one. Then, the impact of the penalty terms will increase, which is not considered in the conventional long packet transmission scheme. We can also find from this figure that a lower value of the decoding error probability requirement brings a lower WST. This is because when $\delta_k$ is large, the value of $Q^{-1}(\delta_k)$ increases, thus leading to larger values of the penalty terms.
\begin{figure}
\vspace{-0.3cm}
\begin{minipage}[t]{0.5\linewidth}
\setcaptionwidth{\textwidth}
\captionsetup{font=small}
\centering
\includegraphics[width=\linewidth]{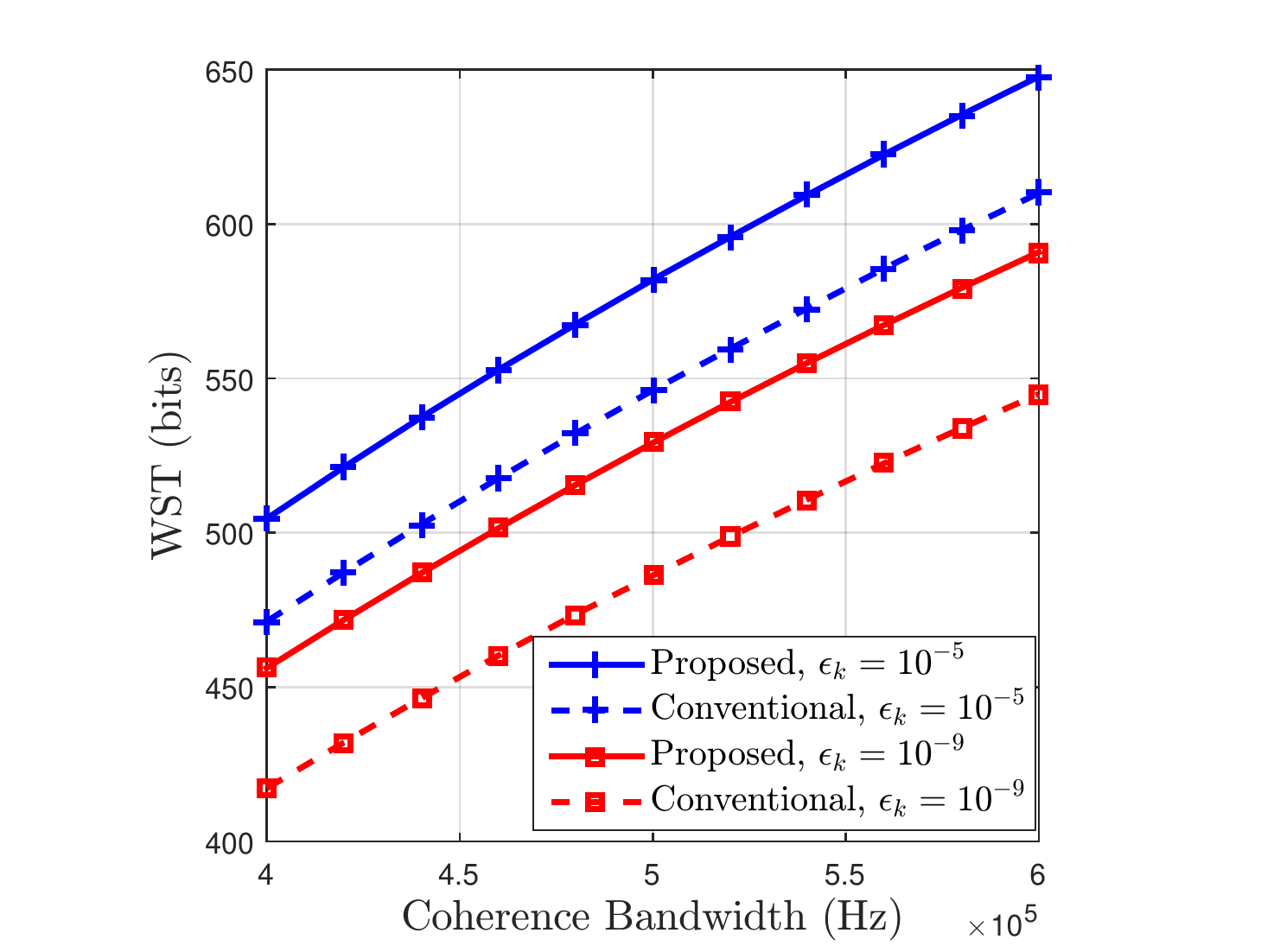}
\vspace{-0.3cm}
\caption{WST versus the channel coherence bandwidth.}
\label{throuvsband}
\vspace{-0.4cm}
\end{minipage}
\begin{minipage}[t]{0.5\linewidth}
\setcaptionwidth{\textwidth}
\captionsetup{font=small}
\centering
\includegraphics[width=\linewidth]{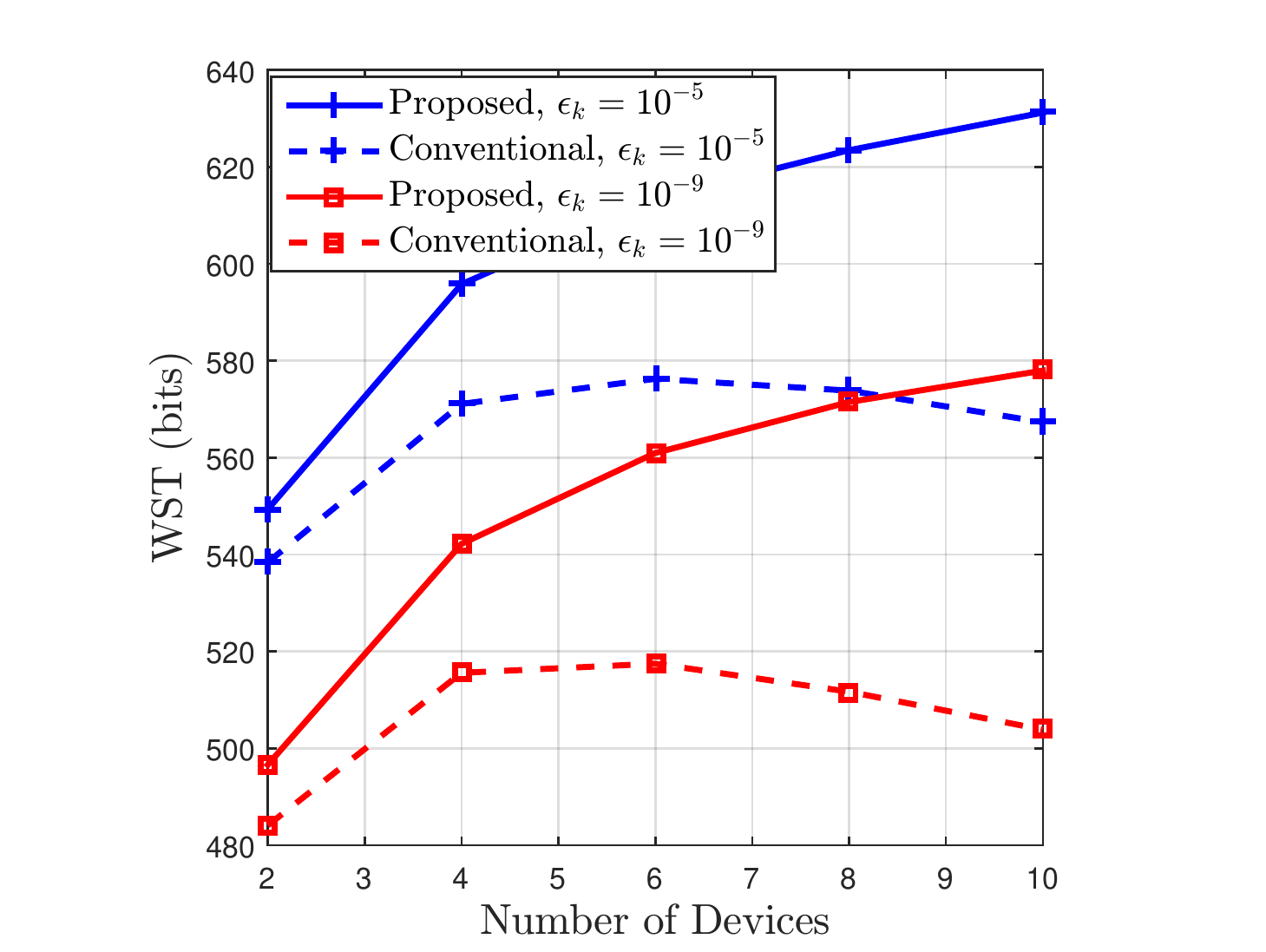}
\vspace{-0.3cm}
\caption{WST versus the number of devices.}
\vspace{-0.4cm}
\label{throuvsK}
\end{minipage}
\end{figure}

Fig.~\ref{throuvsband} shows the WST versus the channel coherence bandwidth $W_c$. In general, higher driving speeds lead to  lower channel coherence bandwidth. We observe from Fig.~\ref{throuvsband} that the WST achieved by all the schemes increase with the increase of channel coherence bandwidth. The reason is that higher channel coherence bandwidth corresponds to larger number of channel uses for each transmission, and thus brings higher throughput. In contrast to Fig.~\ref{throuvspower} where the WST logarithmically increase with increasing transmit power, the WST linearly increases with $W_c$, which demonstrates the significant impact of the channel coherence bandwidth on WST performance. It is again observed that the performance of the proposed BCD algorithm is better than the conventional long packet transmission scheme.

Fig.~\ref{throuvsK} shows the WST versus the number of devices. It is found from this figure that the WST achieved by the proposed BCD algorithm increases with the number of devices as we can employ the multiuser diversity to achieve higher performance. In contrast, the WST of the conventional long packet scheme first increases with $K$ and then decreases with $K$. The main reason is that the conventional long packet scheme targets at optimizing (\ref{dwsfr}) without considering the penalty incurred due to the short packet transmission. The solution that maximizes (\ref{dwsfr}) may not perform well for the short packet throughput formula in (\ref{gedwjij}). This again emphasizes the importance of optimizing the short packet throughput formula in URLLC applications.

\subsection{Total Transmit Power Minimization}
In this subsection, we consider the performance of the proposed method in Section \ref{ttpmin} for the TTP minimization problem. The distance between the AP and the devices are set as: $l_k=100+5(k-1)$ (m), where $k$ denotes the device index. The minimum date packet size is $D_k^{\min}=160$ bits. Three methods are compared. The first one is the solution obtained by solving Problem $({\bf{P4-a}})$ (with legend `Continuous Relaxation'), which is a relaxed version of the original Problem $({\bf{P4}})$. The second one is the solution obtained by converting the continuous solution of $\bf{\bar  N}$ into the discrete solution by using the greedy method (with legend `Integer Conversion'). The final one is the solution obtained by equally allocating the channel bandwidth units to the devices, $n_k=n_{\max}/K$ (with legend `Equal BU Allocation'), and the power allocated to each device can be obtained based on (\ref{yhufyk}).
\begin{figure}
\vspace{-0.3cm}
\begin{minipage}[t]{0.5\linewidth}
\setcaptionwidth{\textwidth}
\captionsetup{font=small}
\centering
\includegraphics[width=\linewidth]{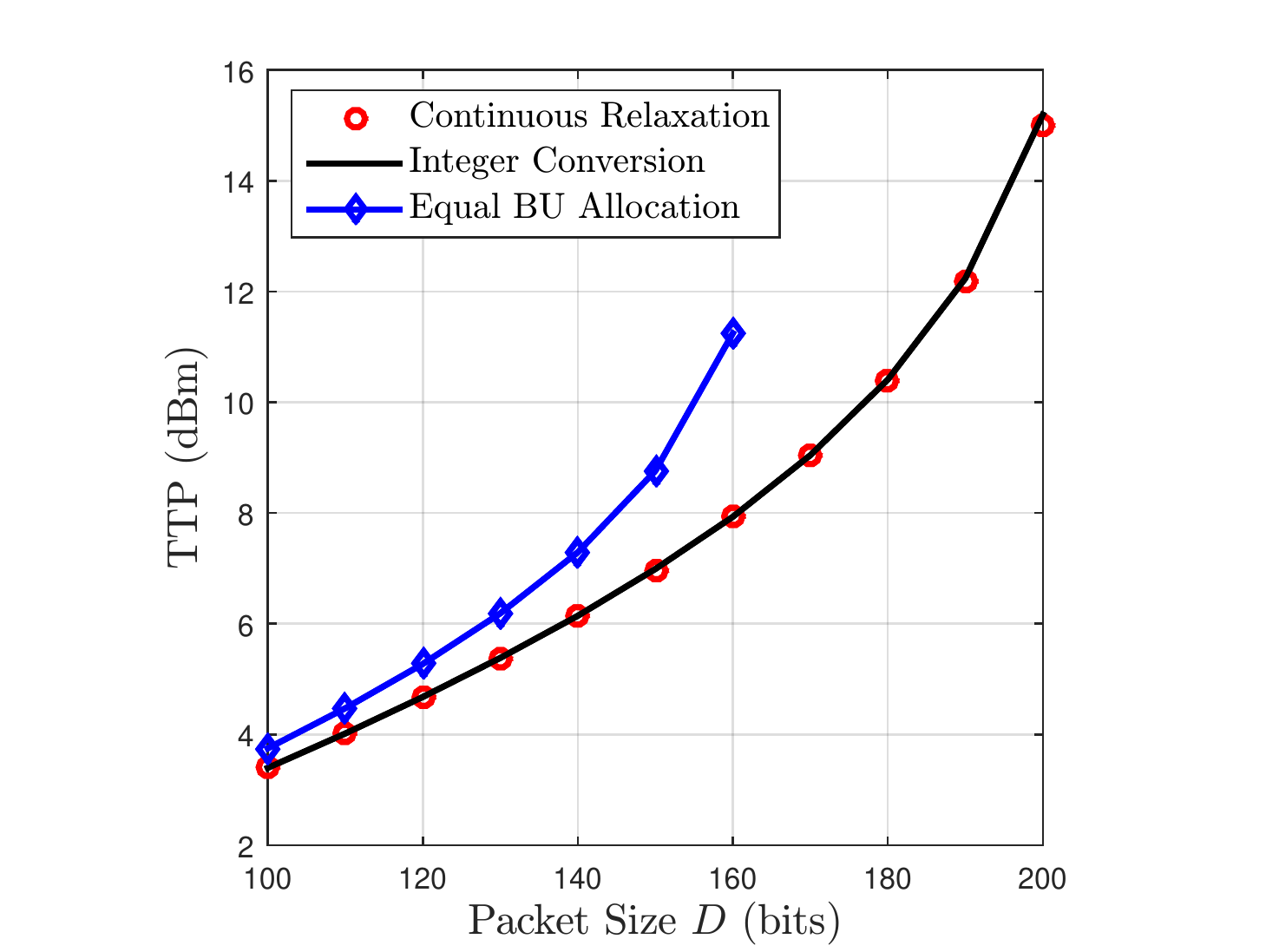}
\vspace{-0.3cm}
\caption{TTP versus the minimum packet size requirement $D$.}
\label{vsd}
\vspace{-0.4cm}
\end{minipage}
\begin{minipage}[t]{0.5\linewidth}
\setcaptionwidth{\textwidth}
\captionsetup{font=small}
\centering
\includegraphics[width=\linewidth]{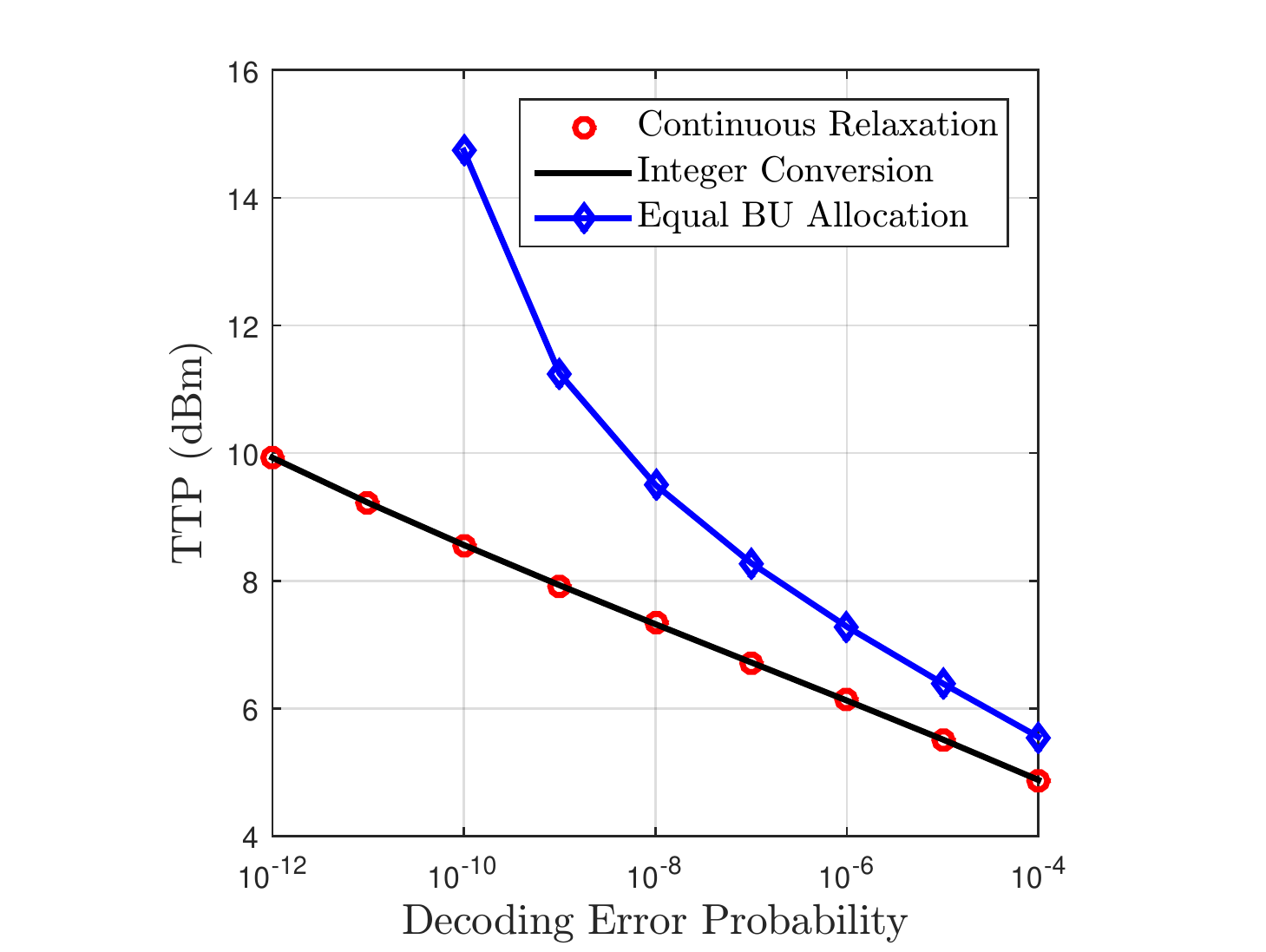}
\vspace{-0.3cm}
\caption{TTP versus the decoding error probability of the devices.}
\vspace{-0.4cm}
\label{vsdec}
\end{minipage}
\end{figure}

\begin{figure}
\vspace{-0.3cm}
\begin{minipage}[t]{0.5\linewidth}
\setcaptionwidth{\textwidth}
\captionsetup{font=small}
\centering
\includegraphics[width=\linewidth]{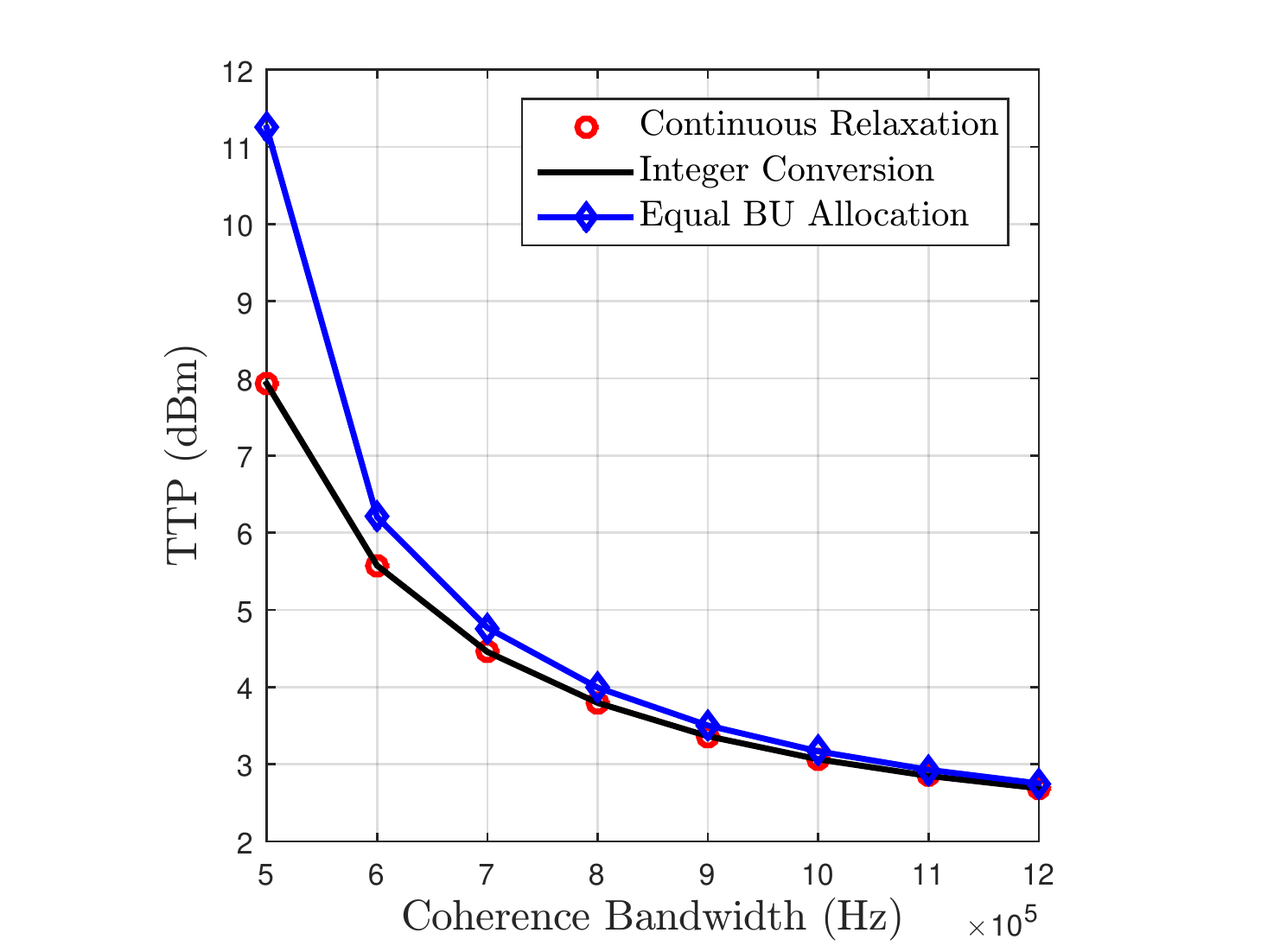}
\vspace{-0.3cm}
\caption{Sum power versus the channel coherence bandwidth.}
\label{vswc}
\vspace{-0.4cm}
\end{minipage}
\begin{minipage}[t]{0.5\linewidth}
\setcaptionwidth{\textwidth}
\captionsetup{font=small}
\centering
\includegraphics[width=\linewidth]{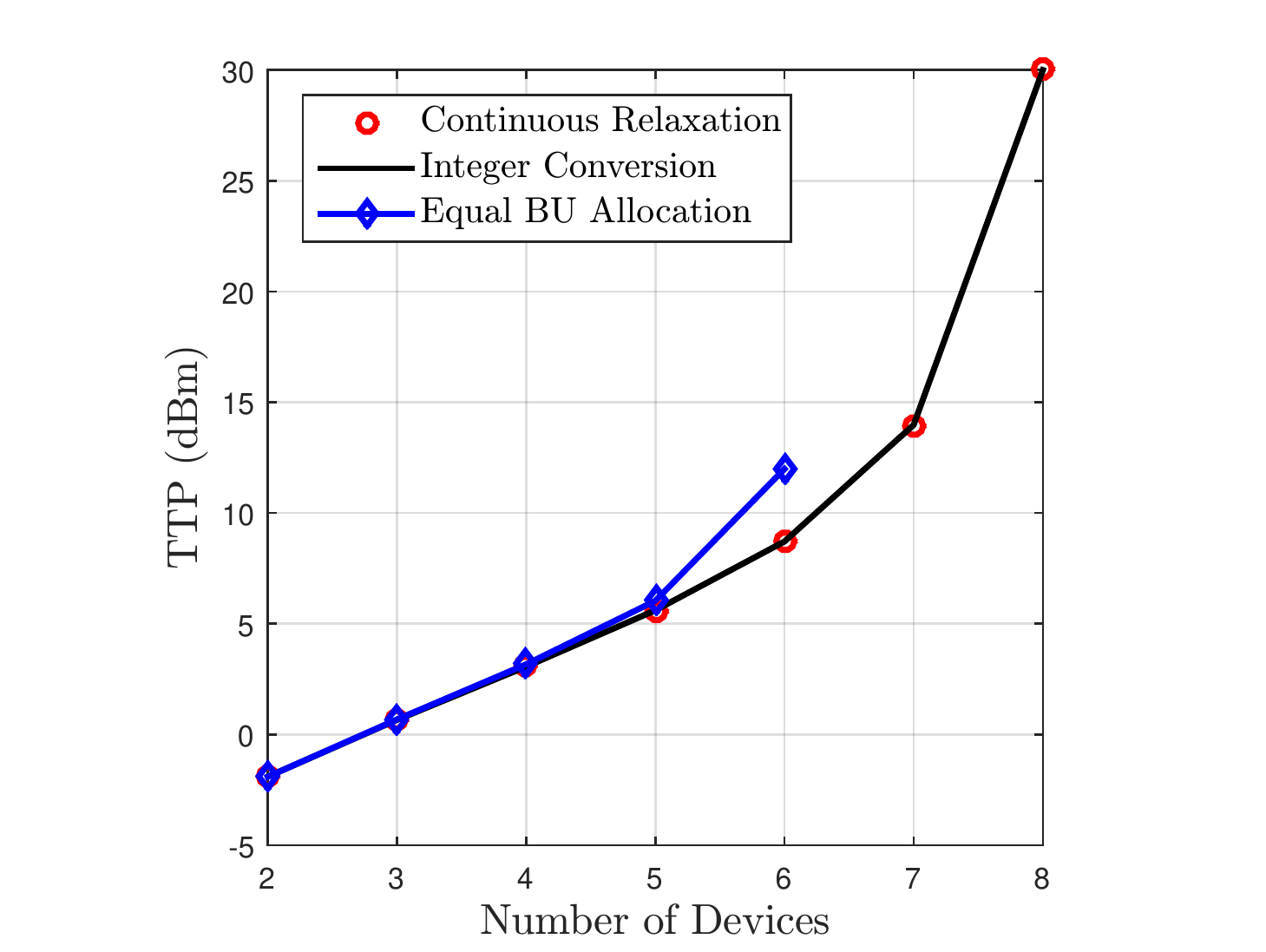}
\vspace{-0.3cm}
\caption{Sum power versus the number of devices.}
\vspace{-0.4cm}
\label{vsnum}
\end{minipage}
\end{figure}

In Fig.~\ref{vsd}, we first study the impact of packet size requirement of each device on the TTP. As  expected, the TTP required monotonically increases with $D$ for all the methods. Both the Continuous Relaxation and the Integer Conversion methods have almost the same performance, which implies the marginal performance loss incurred by the greedy integer conversion procedure. Moreover, both these methods are shown to achieve superior performance over the naive Equal BU Allocation method, and the performance gain monotonically increases with $D$. When $D\ge 160$ bits, the Equal BU Allocation method becomes even infeasible, while the proposed algorithm can support the packet size up to 200 bits. This implies the importance of optimizing the bandwidth unit allocation.

In Fig.~\ref{vsdec}, we investigate the impact of the decoding error probability requirement of the devices on the TTP. It is observed that the TTP required by all the methods decrease with $\epsilon_k$. This can be explained as follows. According to (\ref{reteexpression}), $R_k$ is a monotonically decreasing function of $\epsilon_k$. When $\epsilon_k$ is small, more power is required to achieve the desired data amount requirement. Again, the proposed algorithms are observed to have better performance than the Equal BU Allocation method, especially when the decoding error probability is extremely small.

The impact of the channel bandwidth on the system performance is shown in Fig.~\ref{vswc}. We can find from Fig.~\ref{vswc} that the TTP required by all the methods decreases with increasing channel coherence bandwidth. This is mainly due to the fact that when the channel coherence bandwidth increases, the total number of channel users increases, which can enhance throughput. It is interesting to observe that when the channel coherence bandwidth is sufficiently large, the proposed algorithms can only achieve negligible performance advantage over the Equal BU Allocation method, which implies the equal bandwidth unit allocation is nearly optimal for large channel coherence bandwidth.

Finally, in Fig.~\ref{vsnum}, we study the impact of the number of devices on the system performance where the channel coherence bandwidth is assumed to be $W_c=1$ MHz. It is observed that the sum power increases rapidly with the number of devices. This is because when the number of devices is large, the number of bandwidth units allocated to each device will decrease, which requires more power to transmit the targeted data amount.

\section{Conclusions}\label{jjitrj}
In this paper, we studied a secure mission-critical IoT communication system under URLLC requirements, where the AP transmits safety-critical messages to the devices and there exists an eavesdropper that attempts to eavesdrop this critical message.
Under this context, we considered the WST maximization problem and the TTP minimization problem through joint bandwidth unit and power allocation. For the WST maximization problem, we provided the BCD algorithm to decouple the original coupled optimization problem, and obtain its solution in an iterative manner. For the TTP minimization problem, we derived the sufficient condition when this problem is a convex problem, and we showed that most of the typical URLLC applications satisfy this condition. Low-complexity and efficient algorithms were proposed to find the globally optimal solution, and the greedy method was utilized to convert the continuous solutions into discrete solutions. Simulation results demonstrate the rapid convergence of the BCD algorithm, and performance advantages over the conventional long packet transmission scheme. For the method to solve the TTP minimization problem, simulation results validate the performance advantages in terms of power savings  compared with the naive equal bandwidth unit allocation scheme.

\begin{appendices}
\section{Proof Lemma 1}\label{lemma1}
\bigskip
We first prove that ${f_k({p_k})}$ is a concave function. The second-order derivative of ${f_k({p_k})}$ w.r.t. $p_k$ is given by
\begin{equation}\label{sdefreg}
  f_k''({p_k}) = \frac{{{N_k}}}{{\ln 2}}\frac{{\bar g_k^e - \bar g_k^d}}{{\left( {1 + {p_k}\bar g_k^d} \right)\left( {1 + {p_k}\bar g_k^e} \right)}}\left( {\frac{{\bar g_k^e}}{{1 + {p_k}\bar g_k^e}} + \frac{{\bar g_k^d}}{{1 + {p_k}\bar g_k^d}}} \right) < 0,
\end{equation}
where the last inequality holds since $\bar g_k^d>\bar g_k^e$. Hence, ${f_k({p_k})}$ is a concave function w.r.t. $p_k$. Similarly, second-order derivative of ${y_k({p_k})}$ w.r.t. $p_k$ is given by
\begin{equation}\label{wqdef}
 y_k''({p_k}) =  - \frac{{{{\left( {\bar g_k^d} \right)}^2}{L_k^d}\left( {3 - 2{{\left( {1 + {p_k}\bar g_k^d} \right)}^{ - 2}}} \right)}}{{{{\left( {1 - {{\left( {1 + {p_k}\bar g_k^d} \right)}^{ - 2}}} \right)}^{\frac{3}{2}}}{{\left( {1 + {p_k}\bar g_k^d} \right)}^4}}} - \frac{{{{\left( {\bar g_k^e} \right)}^2}{L_k^e}\left( {3 - 2{{\left( {1 + {p_k}\bar g_k^e} \right)}^{ - 2}}} \right)}}{{{{\left( {1 - {{\left( {1 + {p_k}\bar g_k^e} \right)}^{ - 2}}} \right)}^{\frac{3}{2}}}{{\left( {1 + {p_k}\bar g_k^e} \right)}^4}}}<0.
\end{equation}
Hence, ${y_k({p_k})}$ is a concave function w.r.t. $p_k$.  As a result, $R_k(p_k)$ is the difference of two concave functions $f_k(p_k)$ and $y_k(p_k)$, which completes the proof.

\section{Proof Theorem 1}\label{Theorem1}
Let us define $\lambda\ge 0$ and ${\bm{\mu}}=\{\mu_1,\cdots,\mu_K\}$ as the non-negative dual variables associated with the total power constraint (\ref{powerlimit}) and the individual non-negative power constraint (\ref{nonegp}), respectively. The Lagrangian function of Problem $({\bf{P2-1-a}})$ can be formulated as
\begin{equation}
\mathcal{L} ({\bm{p}},{\bm{\mu}},\lambda)=\sum\nolimits_{k = 1}^K {\left( {\omega _kf_k\left( {{p_k}} \right) - \omega _k{\beta _k}(p_k^{(i - 1)}){p_k}} \right)} - \lambda \left( {\sum\nolimits_{k = 1}^K {{p_k}}  - {P_{\max }}} \right) + \sum\nolimits_{k = 1}^K {{\mu _k}{p_k}}.
\end{equation}
Since Problem $({\bf{P2-1-a}})$  is a convex optimization problem, the globally optimal solution satisfies the Karush-Kuhn-Tucker (KKT) conditions as follows:
\begin{equation}\label{sdefvb}
 \begin{array}{l}
\frac{{\partial \mathcal{L} ({\bm{p}},{\bm{\mu}},\lambda)}}{{\partial {p_k}}} = \frac{{{N_k}}}{{\ln 2}}\frac{{\omega _k(\bar g_k^d - \bar g_k^e)}}{{\left( {1 + {p_k}\bar g_k^d} \right)\left( {1 + {p_k}\bar g_k^e} \right)}} - \omega _k{\beta _k}(p_k^{(i - 1)}) - \lambda  + {\mu _k} = 0,\forall k,\\
{\mu _k}{p_k} = 0,{p_k} \ge 0,\forall k, \lambda \left( {\sum\nolimits_{k = 1}^K {{p_k}}  - {P_{\max }}} \right) = 0,\sum\nolimits_{k = 1}^K {{p_k}}  \le {P_{\max }}.
\end{array}
\end{equation}
Note that $\mu_k,\forall k$ are slack variables in the first equation, which can be eliminated. We then have:
\begin{equation}\label{asvgt}
  \begin{array}{l}
\left( {\omega _k{\beta _k}(p_k^{(i - 1)}) + \lambda  - \frac{{{N_k}}}{{\ln 2}}\frac{\omega _k({\bar g_k^d - \bar g_k^e})}{{\left( {1 + {p_k}\bar g_k^d} \right)\left( {1 + {p_k}\bar g_k^e} \right)}}} \right){p_k} = 0,\forall k,\\
\omega _k{\beta _k}(p_k^{(i - 1)}) + \lambda  \ge \frac{{{N_k}}}{{\ln 2}}\frac{\omega _k({\bar g_k^d - \bar g_k^e})}{{\left( {1 + {p_k}\bar g_k^d} \right)\left( {1 + {p_k}\bar g_k^e} \right)}},\forall k,\\
\lambda \left( {\sum\nolimits_{k = 1}^K {{p_k}}  - {P_{\max }}} \right) = 0,\sum\nolimits_{k = 1}^K {{p_k}}  \le {P_{\max }},{p_k} \ge 0,\forall k.
\end{array}
\end{equation}
By defining $\eta _k^{(i - 1)}(\lambda)$ in Theorem 1, the KKT conditions in (\ref{asvgt}) can be rewritten as
\begin{equation}\label{dfreg}
 \begin{array}{l}
\left( {\left( {1 + {p_k}\bar g_k^d} \right)\left( {1 + {p_k}\bar g_k^e} \right) - \eta _k^{(i - 1)}(\lambda )} \right){p_k} = 0,\forall k,\\
\left( {1 + {p_k}\bar g_k^d} \right)\left( {1 + {p_k}\bar g_k^e} \right) \ge \eta _k^{(i - 1)}(\lambda ),\forall k,\\
\lambda \left( {\sum\nolimits_{k = 1}^K {{p_k}}  - {P_{\max }}} \right) = 0,\sum\nolimits_{k = 1}^K {{p_k}}  \le {P_{\max }},{p_k} \ge 0,\forall k
\end{array}
\end{equation}

If $\eta _k^{(i - 1)}(\lambda ) > 1$, the conditions in (\ref{dfreg}) hold only when $p_k>0$. This can be proved by using contradiction method. Assume that $p_k=0$. Then, based on the second condition of (\ref{dfreg}), we have  $1\ge \eta _k^{(i - 1)}(\lambda )$, which contradicts the condition of $\eta _k^{(i - 1)}(\lambda ) > 1$. Hence, $p_k>0$ should hold. Then, based on the first condition of (\ref{dfreg}), $p_k$ should satisfy the following equation:
\begin{equation}\label{dfrgth}
  {\left( {1 + {p_k}\bar g_k^d} \right)\left( {1 + {p_k}\bar g_k^e} \right) - \eta _k^{(i - 1)}(\lambda )}=0,
\end{equation}
and its solution is given in (\ref{defrfe}) in Theorem 1.

On the other hand, if $\eta _k^{(i - 1)}(\lambda )\le 1$, then $p_k$ must be equal to zero. This can also be proved by using contradiction method. Assume that $p_k>0$. Then, based on the first condition of (\ref{dfreg}), the equation in (\ref{dfrgth}) should hold, and $p_k$ is derived as:
\[{p_k} = \frac{{ - \left( {\bar g_k^d + \bar g_k^e} \right) + \sqrt {{{\left( {\bar g_k^d + \bar g_k^e} \right)}^2} - 4\bar g_k^d\bar g_k^e\left( {1 - \eta _k^{(i - 1)}} \right)} }}{{2\bar g_k^d\bar g_k^e}} \le \frac{{ - \left( {\bar g_k^d + \bar g_k^e} \right) + \sqrt {{{\left( {\bar g_k^d + \bar g_k^e} \right)}^2}} }}{{2\bar g_k^d\bar g_k^e}} = 0,\]
which contradicts the assumption of $p_k>0$. Hence, $p_k$ must be equal to zero.

Combining the above two cases, the optimal solution of $p_k$ is given in (\ref{defrfe}) in Theorem 1. The remaining part of Theorem 1 can be readily proved by using similar analysis for the total power constraint, details of which are omitted for simplicity.

\section{Proof of Lemma 2}\label{lemma2}
We first prove that ${F_k({n_k})}$ is a concave function w.r.t. $n_k$. With some manipulations, the second-order derivative of ${F({n_k})}$ w.r.t. $n_k$ is given by
\begin{equation}\label{xsdwfreag}
F_k''({n_k})= \frac{{{{\tilde N}_0}\left( {\tilde g_k^e - \tilde g_k^d} \right)\left( {\left( {\tilde g_k^e + \tilde g_k^d} \right){n_k} + 2\tilde g_k^e\tilde g_k^d} \right)}}{{{{\ln2\left( {{n_k} + \tilde g_k^e} \right)}^2}{{\left( {{n_k} + \tilde g_k^d} \right)}^2}}} < 0,
\end{equation}
where the inequality holds since ${\bar g_k^e < \bar g_k^d}$. Hence, ${F_k({n_k})}$ is a concave function w.r.t. $n_k$.

 Now, we start to prove that ${G_k({n_k})}$ is also a concave function w.r.t. $n_k$. With some manipulations, the second-order derivative of ${G_k({n_k})}$ w.r.t. $n_k$ is given by
 \begin{equation}\label{defgrthyth}
   G_k''({n_k}) = \frac{{2\frac{{{\partial ^2}z_k^d({n_k})}}{{\partial n_k^2}}z_k^d({n_k}) - {{\left( {\frac{{\partial z_k^d({n_k})}}{{\partial {n_k}}}} \right)}^2}}}{{4z_k^d({n_k})\sqrt {z_k^d({n_k})} }}{{\tilde L}_k^d} + \frac{{2\frac{{{\partial ^2}z_k^e({n_k})}}{{\partial n_k^2}}z_k^e({n_k}) - {{\left( {\frac{{\partial z_k^e({n_k})}}{{\partial {n_k}}}} \right)}^2}}}{{4z_k^e({n_k})\sqrt {z_k^e({n_k})} }}{{\tilde L}_k^e},
 \end{equation}
 where ${\frac{{{\partial ^2}z_k^x({n_k})}}{{\partial n_k^2}}}$ is given by
 \begin{equation}\label{degrt}
   \frac{{{\partial ^2}z_k^x({n_k})}}{{\partial n_k^2}} =  - \frac{{6{n_k}{{\left( {\tilde g_k^x} \right)}^2}}}{{{{\left( {{n_k} + \tilde g_k^x} \right)}^4}}} < 0, x\in \{d,e\}.
 \end{equation}
 Then, combining (\ref{degrt}) and (\ref{defgrthyth}), we know that $ \frac{{{\partial ^2}G_k({n_k})}}{{\partial n_k^2}} <0$. Hence, ${G_k({n_k})}$ is also a concave function w.r.t. $n_k$, and $R_k(n_k)$ is the difference of two concave functions ${F_k({n_k})}$ and ${G_k({n_k})}$.

\section{Proof of Lemma 3}\label{lemma3}
We consider a pair of dual variables $\mu_1$ and $\mu_2$, where $\mu_1\ge \mu_2$. Denote $n_k^\star(\mu_1)$ and $n_k^\star(\mu_2)$ as the optimal solution of Problem (\ref{dwreg}) when $\mu=\mu_1$ and $\mu=\mu_2$, respectively.
Since $n_k^\star(\mu_1)$ is the optimal solution of Problem (\ref{dwreg}) when $\mu=\mu_1$, we have
\begin{equation}\label{dfrgtr}
\begin{array}{l}
{{\cal L}_k}({n_k^\star}({\mu _1}),{\mu _1}) = {F_k}({n_k^\star}({\mu _1})) - {\alpha _k}(n_k^{(j - 1)}){n_k^\star}({\mu _1}) - {\mu _1}{n_k^\star}({\mu _1})\\
 \ge {{\cal L}_k}({n_k^\star}({\mu _2}),{\mu _1}) = {F_k}({n_k^\star}({\mu _2})) - {\alpha _k}(n_k^{(j - 1)}){n_k^\star}({\mu _2}) - {\mu _1}{n_k^\star}({\mu _2}).
\end{array}
\end{equation}
Furthermore, $n_k^\star(\mu_2)$ is the optimal solution of Problem (\ref{dwreg}) when $\mu=\mu_2$, we have
\begin{equation}\label{dergthy}
 \begin{array}{l}
{{\cal L}_k}({n_k^\star}({\mu _2}),{\mu _2}) = {F_k}({n_k^\star}({\mu _2})) - {\alpha _k}(n_k^{(j - 1)}){n_k^\star}({\mu _2}) - {\mu _2}{n_k^\star}({\mu _2})\\
 \ge {{\cal L}_k}({n_k^\star}({\mu _1}),{\mu _2}) = {F_k}({n_k^\star}({\mu _1})) - {\alpha _k}(n_k^{(j - 1)}){n_k^\star}({\mu _1}) - {\mu _2}{n_k^\star}({\mu _1}).
\end{array}
\end{equation}
By adding these two inequalities and simplifying them, we have $\left( {{n_k^\star}({\mu _1}) - {n_k^\star}({\mu _2})} \right)\left( {{\mu _2} - {\mu _1}} \right) \ge 0$. Since $\mu_1>\mu_2$, we have ${n_k^\star}({\mu _1}) < {n_k^\star}({\mu _2})$.
Then, by summing all these $K$ inequalities, we have $W(\mu_1)=\sum\nolimits_{k = 1}^K {{n_k^\star}({\mu _1})}<\sum\nolimits_{k = 1}^K {{n_k^\star}({\mu _2})}=W(\mu _2)$. Hence, $W(\mu)$ is a monotonically decreasing function of $\mu$.

\section{Proof of Theorem 2}\label{theorem2}

We first prove its convexity. With some manipulations, the second-order derivative of ${p_k}({N_k})$ w.r.t. $N_k$ is calculated as
\begin{equation}\label{vfeagtr}
 {{p''_k}}({N_k})=\frac{{\frac{{{c_k}{d_k}}}{{N_k^3}}{e^{\frac{{{a_k}}}{{{N_k}}} + \frac{{{b_k}}}{{\sqrt {{N_k}} }}}}\Xi ({N_k}) + \frac{{{c_k}}}{{N_k^3}}{e^{\frac{{2{a_k}}}{{{N_k}}} + \frac{{2{b_k}}}{{\sqrt {{N_k}} }}}}\Phi ({N_k})}}{{{{\left( {{d_k} - {e^{\frac{{{a_k}}}{{{N_k}}} + \frac{{{b_k}}}{{\sqrt {{N_k}} }}}}} \right)}^3}}},
\end{equation}
where $\Xi ({N_k})$ and $\Phi ({N_k})$ are given by
\begin{equation}\label{adaqd}
  \Xi ({N_k}) =  - \frac{{{b_k}}}{4}N_k^{\frac{3}{2}} + \frac{{b_k^2}}{4}{N_k} + {a_k}{b_k}N_k^{\frac{1}{2}} + a_k^2, \Phi ({N_k}) = \frac{{{b_k}}}{4}N_k^{\frac{3}{2}} + \frac{{b_k^2}}{4}{N_k} + {a_k}{b_k}N_k^{\frac{1}{2}} + a_k^2.
\end{equation}
 Since ${N_k}>N_k^{{\rm{lb}}}$, the denominator of (\ref{vfeagtr}) is larger than zero. Obviously, $\Phi ({N_k})$ is larger than zero. Hence, if $\Xi ({N_k})>0$, then ${{p''_k}}({N_k})>0$ holds and ${{p_k}}({N_k})$ is a convex function of $N_k$. Next, we derive the condition when $\Xi ({N_k})>0$.

 Denote $t_k = N_k^{\frac{1}{2}}$. Then, $\Xi ({N_k}) $ can be re-expressed as
 \begin{equation}\label{csfcrfr}
\Xi ({t_k}) =  -\frac{{{b_k}}}{4}t_k^3 + \frac{{b_k^2}}{4}t_k^2 + {a_k}{b_k}{t_k} + a_k^2.
 \end{equation}
 Note that $\Xi (0)=a_k^2>0$ and $\Xi (+\infty )=-\infty $. Since $\Xi ({t_k})$ is a continuous function, there must exist at least one positive solution for the equation $\Xi ({t_k})=0$. In the following, we prove that the solution is unique.

We rewrite equation $\Xi ({t_k})=0$ as a standard cubic equation:
\begin{equation}\label{johpipkww}
u_k{t_k^3} + v_k{t_k^2} + w_kt_k + z_k = 0,
\end{equation}
where $u_k=-\frac{{{b_k}}}{4}$, $v_k=\frac{{b_k^2}}{4}$, $w_k={a_k}{b_k}$, and $z_k=a_k^2$.

By dividing (\ref{johpipkww}) by $u_k$ and inserting $t_k=x_k-v_k/3u_k$, we have
\begin{equation}\label{swdeartrsg}
{x_k^3} + \rho_k x_k+ \kappa_k  = 0,
\end{equation}
where $\rho_k$ and $\kappa_k$  are defined in Theorem 2. It can be readily verified that
\begin{equation}\label{vfeafae}
 4{\rho_k^3} + 27{\kappa_k ^2} > 0, \kappa_k>0, \rho_k  < 0.
\end{equation}
As a result, there exists only one real solution for (\ref{swdeartrsg}), which is given by
\begin{equation}\label{jojasdswetjhy}
 x_k^* =  - 2 \sqrt { - \frac{\rho_k }{3}} \cosh \left( {\frac{1}{3}{\rm{arcosh}}\left( {\frac{{ - 3\kappa_k}}{{2\rho_k }}\sqrt {\frac{{ - 3}}{\rho_k }} } \right)} \right).
\end{equation}
Thus, the unique solution of equation (\ref{johpipkww}) is given by $t_k^*=x_k^*-v_k/3u_k$.

Based on the above discussion, we can conclude that when $t_k<t_k^*=x_k^*-v_k/3u_k$, $\Xi ({t_k})$ is positive and ${{p_k}}({N_k})$ is a convex function.

Now we proceed to prove that ${{p_k}}({N_k})$ is a monotonically decreasing function when inequality (\ref{fdvtgxdtr}) holds. The first-order derivative of ${{p_k}}({N_k})$ w.r.t. $p_k$ is given by
\begin{equation}\label{adcwefdwef}
  {{p'_k}}({N_k}) =  - \frac{1}{{{{\tilde h}^e}}} + \frac{{{c_k}\left( {{d_k} - {e^{\frac{{{a_k}}}{{{N_k}}} + \frac{{{b_k}}}{{\sqrt {{N_k}} }}}}} \right) - {c_k}{e^{\frac{{{a_k}}}{{{N_k}}} + \frac{{{b_k}}}{{\sqrt {{N_k}} }}}}\left( {\frac{{{a_k}}}{{{N_k}}} + \frac{{{b_k}}}{{\sqrt {{N_k}} }}} \right)}}{{{{\left( {{d_k} - {e^{\frac{{{a_k}}}{{{N_k}}} + \frac{{{b_k}}}{{\sqrt {{N_k}} }}}}} \right)}^2}}}.
\end{equation}
Since ${{p_k}}({N_k})$ is a convex function when inequality (\ref{fdvtgxdtr}) holds, ${{p'_k}}({N_k})$ is a monotonically increasing function. Then, we have
\begin{equation}\label{dcefra}
\vspace{-0.4cm}
{{p'_k}}({N_k}) < {{p'_k}}(\infty ) = 0.
\end{equation}
In consequence, ${{p_k}}({N_k})$ is a monotonically deceasing function when inequality (\ref{fdvtgxdtr}) holds.
\end{appendices}

\bibliographystyle{IEEEtran}
\bibliography{myre}

\begin{thebibliography}{10}
\providecommand{\url}[1]{#1}
\csname url@samestyle\endcsname
\providecommand{\newblock}{\relax}
\providecommand{\bibinfo}[2]{#2}
\providecommand{\BIBentrySTDinterwordspacing}{\spaceskip=0pt\relax}
\providecommand{\BIBentryALTinterwordstretchfactor}{4}
\providecommand{\BIBentryALTinterwordspacing}{\spaceskip=\fontdimen2\font plus
\BIBentryALTinterwordstretchfactor\fontdimen3\font minus
  \fontdimen4\font\relax}
\providecommand{\BIBforeignlanguage}[2]{{%
\expandafter\ifx\csname l@#1\endcsname\relax
\typeout{** WARNING: IEEEtran.bst: No hyphenation pattern has been}%
\typeout{** loaded for the language `#1'. Using the pattern for}%
\typeout{** the default language instead.}%
\else
\language=\csname l@#1\endcsname
\fi
#2}}
\providecommand{\BIBdecl}{\relax}
\BIBdecl

\bibitem{Shafi2017}
M.~{Shafi}, A.~F. {Molisch}, P.~J. {Smith}, T.~{Haustein}, P.~{Zhu}, P.~{De
  Silva}, F.~{Tufvesson}, A.~{Benjebbour}, and G.~{Wunder}, ``{5G}: A tutorial
  overview of standards, trials, challenges, deployment, and practice,''
  \emph{IEEE J. Sel. Areas Commun.}, vol.~35, no.~6, pp. 1201--1221, June 2017.

\bibitem{Polyanskiy2010}
Y.~Polyanskiy, H.~V. Poor, and S.~Verdu, ``Channel coding rate in the finite
  blocklength regime,'' \emph{IEEE Trans. Inf. Theory}, vol.~56, no.~5, pp.
  2307--2359, May 2010.

\bibitem{chen2019resource}
J.~Chen, L.~Zhang, Y.-C. Liang, X.~Kang, and R.~Zhang, ``Resource allocation
  for wireless-powered {IoT} networks with short packet communication,''
  \emph{IEEE Trans. Wireless Commun.}, vol.~18, no.~2, pp. 1447--1461, 2019.

\bibitem{pan2019joint}
C.~Pan, H.~Ren, Y.~Deng, M.~Elkashlan, and A.~Nallanathan, ``Joint blocklength
  and location optimization for {URLLC}-enabled {UAV} relay systems,''
  \emph{IEEE Commun. Lett.}, vol.~23, no.~3, pp. 498--501, March.

\bibitem{hongren2019wcl}
H.~{Ren}, C.~{Pan}, K.~{Wang}, Y.~{Deng}, M.~{Elkashlan}, and A.~{Nallanathan},
  ``Achievable data rate for {URLLC}-enabled {UAV} systems with {3-D} channel
  model,'' \emph{IEEE Wireless Commun. Lett.}, pp. 1--1, 2019.

\bibitem{sun2018short}
X.~Sun, S.~Yan, N.~Yang, Z.~Ding, C.~Shen, and Z.~Zhong, ``Short-packet
  downlink transmission with non-orthogonal multiple access,'' \emph{IEEE
  Trans. Wireless Commun.}, vol.~17, no.~7, pp. 4550--4564, 2018.

\bibitem{she2019cross}
C.~She, Y.~Duan, G.~Zhao, T.~Q. Quek, Y.~Li, and B.~Vucetic, ``Cross-layer
  design for mission-critical {IoT} in mobile edge computing systems,''
  \emph{IEEE Internet Things J.}, 2019.

\bibitem{hong-twc}
H.~{Ren}, C.~{Pan}, Y.~{Deng}, M.~{Elkashlan}, and A.~{Nallanathan}, ``Joint
  power and blocklength optimization for {URLLC} in a factory automation
  scenario,'' \emph{IEEE Trans. Wireless Commun.}, pp. 1--1, 2019.

\bibitem{hong-jasc}
------, ``Joint pilot and payload power allocation for {Massive-MIMO-enabled
  URLLC IIoT} networks,'' \emph{to appear in IEEE J. Sel. Areas Commun.},
  https://arxiv.org/abs/1912.12438.

\bibitem{changyang2018}
C.~She, C.~Yang, and T.~Q.~S. Quek, ``Joint uplink and downlink resource
  configuration for ultra-reliable and low-latency communications,'' \emph{IEEE
  Trans. Commun.}, vol.~66, no.~5, pp. 2266--2280, May 2018.

\bibitem{mukherjee2014principles}
A.~Mukherjee, S.~A.~A. Fakoorian, J.~Huang, and A.~L. Swindlehurst,
  ``Principles of physical layer security in multiuser wireless networks: A
  survey,'' \emph{IEEE Commun. Surveys Tuts.}, vol.~16, no.~3, pp. 1550--1573,
  2014.

\bibitem{mukherjee2015physical}
A.~Mukherjee, ``Physical-layer security in the internet of things: Sensing and
  communication confidentiality under resource constraints,'' \emph{Proceedings
  of the IEEE}, vol. 103, no.~10, pp. 1747--1761, 2015.

\bibitem{li2019joint}
Z.~Li, M.~Chen, C.~Pan, N.~Huang, Z.~Yang, and A.~Nallanathan, ``Joint
  trajectory and communication design for secure {UAV} networks,'' \emph{IEEE
  Commun. Letters}, vol.~23, no.~4, pp. 636--639, 2019.

\bibitem{yizhoutcom}
Y.~{Zhou}, C.~{Pan}, P.~L. {Yeoh}, K.~{Wang}, M.~{Elkashlan}, B.~{Vucetic}, and
  Y.~{Li}, ``Secure communications for {UAV}-enabled mobile edge computing
  systems,'' \emph{IEEE Trans. Commun.}, pp. 1--1, 2019.

\bibitem{weiyang-tit-19}
W.~{Yang}, R.~F. {Schaefer}, and H.~V. {Poor}, ``Wiretap channels:
  Nonasymptotic fundamental limits,'' \emph{IEEE Trans. Inf. Theory}, vol.~65,
  no.~7, pp. 4069--4093, July 2019.

\bibitem{huimingwang2019}
H.~{Wang}, Q.~{Yang}, Z.~{Ding}, and H.~V. {Poor}, ``Secure short-packet
  communications for mission-critical {IoT} applications,'' \emph{IEEE Trans.
  Wireless Commun.}, vol.~18, no.~5, pp. 2565--2578, May 2019.

\bibitem{hongrenICC}
H.~{Ren}, C.~{Pan}, Y.~{Deng}, M.~{Elkashlan}, and A.~{Nallanathan}, ``Resource
  allocation for {URLLC} in {5G} mission-critical {IoT} networks,'' in
  \emph{2019 IEEE ICC}, May 2019, pp. 1--6.

\bibitem{dinh2010local}
Q.~T. Dinh and M.~Diehl, ``Local convergence of sequential convex programming
  for nonconvex optimization,'' in \emph{Recent Advances in Optimization and
  its Applications in Engineering}.\hskip 1em plus 0.5em minus 0.4em\relax
  Springer, 2010, pp. 93--102.

\bibitem{boyd2004convex}
S.~Boyd and L.~Vandenberghe, \emph{Convex optimization}.\hskip 1em plus 0.5em
  minus 0.4em\relax Cambridge university press, 2004.

\bibitem{cunhua2019}
C.~{Pan}, H.~{Ren}, M.~{Elkashlan}, A.~{Nallanathan}, and L.~{Hanzo}, ``Robust
  beamforming design for ultra-dense user-centric {C-RAN} in the face of
  realistic pilot contamination and limited feedback,'' \emph{IEEE Trans.
  Wireless Commun.}, vol.~18, no.~2, pp. 780--795, Feb 2019.

\bibitem{changjian}
C.~{Sun}, C.~{She}, C.~{Yang}, T.~Q.~S. {Quek}, Y.~{Li}, and B.~{Vucetic},
  ``Optimizing resource allocation in the short blocklength regime for
  ultra-reliable and low-latency communications,'' \emph{IEEE Trans. Wireless
  Commun.}, vol.~18, no.~1, pp. 402--415, Jan 2019.

\bibitem{access2010further}
E.~U. T.~R. Access, ``Further advancements for {E-UTRA} physical layer
  aspects,'' \emph{3GPP TR 36.814, Tech. Rep.}, 2010.

\end{thebibliography}


\end{document}